\documentclass[]{aastex631}

\usepackage{amssymb}
\usepackage{amsmath}
\usepackage{xcolor}

\newcommand\cp{C$_{\rm{P}}$}
\newcommand\fermi{\textit{Fermi }}
\newcommand\fermilat{\textit{Fermi}-LAT}

\begin{document}

\title{A Cross-correlation Study between IceCube Neutrino Events and the \fermi Unresolved Gamma-ray Sky.}

\author[0000-0002-6548-5622]{Michela Negro}
\correspondingauthor{Michela Negro}
\email{mnegro1@umbc.edu}
\affiliation{University of Maryland, Baltimore County, Baltimore, MD 21250, USA}
\affiliation{NASA Goddard Space Flight Center, Greenbelt, MD 20771, USA}
\affiliation{Center for Research and Exploration in Space Science and Technology, NASA/GSFC, Greenbelt, MD 20771, USA}

\author[0000-0002-7604-1779]{Milena Crnogor\v{c}evi\'{c}}
\affiliation{Center for Research and Exploration in Space Science and Technology, NASA/GSFC, Greenbelt, MD 20771, USA}
\affiliation{Department of Astronomy, University of Maryland, College Park, MD 20742, USA}

% \author[0000-0002-6996-1155]{Michael Larson}
% \affiliation{Department of Physics, University of Maryland, College Park, MD 20742, USA}

% \author[0000-0002-5387-8138]{Ke Fang}
% \affiliation{Department of Physics, Wisconsin IceCube Particle Astrophysics Center, University of Wisconsin, Madison, WI, 53706}

\author[0000-0002-2942-3379]{Eric Burns}
\affiliation{Louisiana State University, Baton Rouge, LA 70803, USA}

\author[0000-0002-3925-7802]{Eric Charles}
\affiliation{Kavli Institute for Particle Astrophysics and Cosmology (KIPAC), Stanford University, Stanford, CA 94305, USA}
\affiliation{SLAC National Accelerator Laboratory, 2575 Sand Hill Road, Menlo Park, CA 94025, USA}
%\collaboration{(LaTeX collaboration)}

\author[0000-0002-8472-3649]{Lea Marcotulli}
\altaffiliation{NHFP Einstein Fellow}
\affil{Yale Center for Astronomy \& Astrophysics, 52 Hillhouse Avenue, New Haven, CT 06511, USA}
\affil{Department of Physics, Yale University, P.O. Box 208120, New Haven, CT 06520, USA}

\author[0000-0002-9280-836X]{Regina Caputo}
\affiliation{NASA Goddard Space Flight Center, Greenbelt, MD 20771, USA}

\begin{abstract}

With the coincident detections of electromagnetic radiation together with gravitational waves (GW170817) or neutrinos (TXS 0506+056), the new era of multimessenger astrophysics has begun. Of particular interest are the searches for correlation between the high-energy astrophysical neutrinos detected by the IceCube Observatory and gamma-ray photons detected by the Fermi Large Area Telescope (LAT). So far, only sources detected by the LAT have been considered in correlation with IceCube neutrinos, neglecting any emission from sources too faint to be resolved individually. Here, we present the first cross-correlation analysis considering the unresolved gamma-ray background (UGRB) and IceCube events. We perform a thorough sensitivity study and, given the lack of identified correlation, we place upper limits on the fraction of the observed neutrinos that would be produced in proton-proton or proton-$\gamma$ interactions from the population of sources contributing to the UGRB emission and dominating its spatial anisotropy (aka blazars). Our analysis suggests that, under the assumption that there is no intrinsic cutoff and/or hardening of the spectrum above \fermilat{} energies, and that all gamma-rays from the unresolved blazars dominating the UGRB fluctuation field are produced by neutral pions from \textit{p-p} (\textit{p-}$\gamma$) interactions, up to 60\% (30\%) of such population may contribute to the total neutrino events observed by IceCube. This translates into a O(1\%) maximum contribution to the astrophysical high-energy neutrino flux observed by IceCube at 100 TeV.

\end{abstract}

\keywords{Gamma-rays - Neutrinos - Multimessenger}

% \tableofcontents

% \clearpage

\section{Introduction} \label{sec:intro}
A population of high-energy neutrinos of astrophysical origin has been observed \citep{2013Sci...342E...1I}, though the sources of the bulk of these events remain unknown \citep{IceCube:2020acn}. 
Neutrinos at TeV-PeV energies are typically produced when relativistic protons interact with matter via hadronuclear interaction (inelastic $p$-$p$ scattering) or with radiation via photohadronic processes ($p$-$\gamma$ interactions). Such processes also produce neutral pions that decay into gamma rays. Since neither signal experiences propagation delay, a simultaneous detection of gamma rays and neutrinos would provide invaluable insights into the nature of the production source.

Blazars are promising sources of high-energy neutrinos. The blazar TXS~0506$+$056 was identified as the first high-energy neutrino source candidate via the coincidence of its very-high-energy gamma-ray flare with a high-energy neutrino \citep{IceCube:2018dnn} and excess of neutrino events in IceCube's historical data \citep{IceCube:2018cha}. In addition, an excess of IceCube neutrinos has been found from the direction of the starburst Seyfert galaxy NGC~1068 \citep{ICM77}, also detected in gamma rays in the GeV regime. Marginal evidence has been suggested toward spatial coincidences between tidal disruption events and IceCube alert events, albeit with a relatively long delay in neutrino detection \citep{PhysRevLett.124.051103, Stein:2020xhk}. These results suggest that the production of high-energy neutrinos could be related to supermassive black hole activity. 

Besides single sources studies in coincidence with IceCube events, significant effort has been directed towards systematic searches of spatial correlation between blazar catalogs and IceCube all-sky data. Searches using gamma-ray blazars, including the second  \fermi Large Area Telescope (LAT) Active Galactic Nuclei (AGN) catalog \citep[2LAC,][]{2LAC} and the {\it Fermi}-LAT low energy catalog \citep[1FLE,][]{1FLE}, found that resolved blazars contribute less than $\sim 30\%$ and 1\% of the diffuse flux \citep{IceCube:2016qvd, IceCube:2022zbd}, respectively. Searches with radio-selected blazars using public IceCube data introduced possible positive spatial correlations \citep{2020ApJ...894..101P, Buson:2022fyf}. 

Although the sample of resolved blazars includes the brightest sources, the cumulative emission from unresolved sources carries substantial amount of energy. In fact, the spatial distribution of the Unresolved Gamma-ray Background (UGRB) sources and its corresponding anisotropy measurement (most recently measured by \cite{UGRBaniso2018}) can entirely be accounted for by the gamma-ray emission from isotropically distributed blazars below the LAT sensitivity threshold \citep{ManconiAnisoInterpr, UGRBanisoInterpr}. At the same time, this very population of blazars accounts for the 20--30\% of the total UGRB intensity spectrum \citep{UGRBanisoInterpr}, which also sees the contribution from other, more numerous and spatially smoother, populations such as star forming galaxies and misaligned AGNs (see, e.g., \cite{SFGLinden} and \cite{IGRBmAGN}). Finally, GeV-TeV photons from extragalactic gamma-ray sources will be attenuated due to pair production, resulting in the brightest neutrino sources being gamma-ray dim \citep{2022ApJ...933..190F}. It is therefore crucial to study the contribution of these faint, unresolved gamma-ray blazars to the diffuse high-energy neutrino flux. 

So far, source association studies of astrophysical neutrinos have largely relied upon considering resolved gamma-ray sources; in turn, neglecting all the objects that are too faint to be detected individually, yet still contributing to the total gamma-ray flux. In this work, we determine the level of correlation between the UGRB emission as observed by the \fermi over 12 years, and the neutrino event observations conducted in 10 years of IceCube observations \citep{TheLAT, IC10yr}.

Previous source association involving IceCube data analyses mostly rely on likelihood stacking (e.g., \citealp{IceCube:2020acn}). More recently, the two-point cross-correlation method has also been employed to investigate the physical origins of astrophysical neutrinos. Cross-correlation measurements were primarily introduced to describe the distribution of galaxies in the Universe, relying upon the consideration of excess probabilities of finding two galaxies at some separation, drawn from a random distribution of points \citep{1980lssu.book.....P}. This cross-correlation technique has been extensively adopted to characterize the UGRB in several works studying its connection with the large scale structure of the Universe: from galaxy catalogs \citep{xia11, Cuoco:2017bpv, Ammazzalorso:2018evf} to galaxy cluster catalogs \citep{Branchini:2016glc, 2017arXiv170809385L, 2017arXiv170900416L, 2018PASJ...70S..25M}, weak lensing from cosmic shear \citep{Camera:2012cj, Camera:2014rja, Shirasaki:2014noa, AmmazzalorsoDES}, and lensing potential of the cosmic-microwave background \citep{Fornengo:2014cya}. A generalization of such method was utilized in \cite{Fang:2020rvq}, considering association of astrophysical neutrinos with the well-calibrated tracers of the large scale structure obtained from the infrared catalogs.

As such, in this paper we investigate the sensitivity of the 2D spatial cross-correlation technique to detecting a significant cross-correlation signal between the emission from a population of LAT-unresolved $\gamma$-ray blazars and the muon-neutrino events detected by IceCube. To this end, we build a simulation pipeline with the goal of comparing the sensitivity by varying the amount of expected observed neutrino signal given a gamma-ray intensity distribution and assuming \textit{p-p} or \textit{p-}$\gamma$ interaction. We also investigate the improvement in sensitivity with increased statistics in the IceCube data sample, in view of future advancements and data reprocessing such as IceCube-Gen2 \citep{ICGen2}. The paper is organized as follows. Section \ref{sec:wherehow} provides the descriptions of the analysis set-up and the cross-correlation method. We devote Section \ref{sec:real} to the construction of the data maps, while a description of the procedure to generate simulated maps is laid out in Section \ref{sec:sim}. The results for both the sensitivity study and the real-data cross-correlation analysis are shown in Section \ref{sec:res}. Finally, the discussion and the conclusions are presented in Section \ref{sec:concl}. Additional considerations and plots are provided in Appendix.

\section{Where to look and how}
\label{sec:wherehow}

In this section we briefly illustrate the technique used to compute the cross-correlation angular power spectrum and then we discuss the main features of the data from the two observatories, \fermilat{} and IceCube, that justify the selections and the analysis set up adopted in this work. 

\subsection{CAPS computation}

The cross-correlation angular power spectrum (CAPS) between a field $\delta_\alpha$ and a field $\delta_\beta$ is defined as
\begin{equation}
    C_\ell^{\alpha\beta} = \frac{1}{2\ell+1}\left< \sum_m a^\alpha_{\ell m}a^\beta_{\ell m}\right>
\end{equation}
where the brackets indicate the average on the modes $m$, and the coefficients 
$a_{\ell m}$ are given by the expansion into spherical harmonics of the fields under study: 
\begin{equation}
\delta_\alpha({\rm{\bf n}}) = \sum_{\ell m} a^\alpha_{\ell m} Y_{\ell m}({\rm{\bf n}})~,
\end{equation}
\noindent
where ${\rm{\bf n}}$ denotes a given direction in the sky. Hence, the CAPS is a measurement of the amplitude of the anisotropy associated to different multipoles, $\ell$, which correspond to different angular scales, $\theta$. Higher multipoles correspond to smaller angular scales.

In our study the two fields are represented by the gamma-ray intensity field (in units of cm$^{-2}$s$^{-1}$sr$^{-1}$) and the neutrino count fluctuation field: 
\begin{equation}
    \delta_\gamma({\rm{\bf n}}) = \Phi_\gamma({\rm{\bf n}})~~~~~~
    \delta_\nu({\rm{\bf n}}) = \Xi_\nu({\rm{\bf n}})
\end{equation}
where $\Xi_\nu({\rm{\bf n}})$ is defined later on in Eq.~\ref{eq:ICfluct}.
While the generation of such field maps in HEALPix format is detailed in Section~\ref{sec:sim} and Section~\ref{sec:real} for simulated and real data respectively, here we describe the procedure to compute the CAPS.
Given a pair of HEALPix\footnote{\url{http://healpix.sourceforge.net}} \citep{HEALPix1, HEALPix2} maps of equal order, the CAPS are computed exploiting the \textit{PolSpice} statistical toolkit \citep{szapudi01, chon04, efstathiou04, challinor05}. \textit{PolSpice} automatically corrects the angular power spectra for the effect of a mask applied to the maps and additionally it provides the covariance matrix, $V_{\ell\ell'}$, which carries information about the covariant uncertainties among the power of different angular scales. 
% We use the covaariance matrix to estimate the error on each C$_{\ell}$. %\textcolor{red}{We defone the best multipole range for our study ....}.

In order to reduce their contamination on the small scales angular power, we remove the monopole and the dipole components by using the dedicated \textit{PolSpice}'s routines, which relies on the HEALPix {\tt remove\_dipole} function. The CAPS, as given by \textit{PolSpice}, must be corrected by the point spread function (PSF) of both \fermilat{} and IceCube. Additionally, a correction must be applied to account for the spatial binning (pixeling) of the maps. Let us define $C_{\ell}^{(\gamma \nu)}$ the raw CAPS, from which we can obtain the corrected CAPS, $\overline{C}_{\ell}^{(\gamma \nu)}$, as: 

\begin{equation}
\overline{C}_{\ell}^{(\gamma \nu)} = C_{\ell}^{(\gamma \nu)}~W_{\ell}^{-1}
\end{equation}
\noindent
where $W_{\ell} = (W_{\ell}^{\gamma,beam}W_{\ell}^{\nu,beam})(W_{\ell}^{pix})^2$. $W_{pix}$ is called ``pixel window function'' and corrects for the spatial binning used to map the events. $W_{\ell}^{\gamma,beam}$ and $W_{\ell}^{\nu,beam}$ are the so-called ``beam window functions'' for \fermi and IceCube respectively, and account for the PSF profiles of the instruments. They are computed as: 

\begin{equation}
W^{beam}(E,\ell) = 2\pi \int_{0}^{\pi} P_{\ell}(\cos\theta) \textrm{PSF}(\theta,E) \, \sin\theta d\theta
\label{eq:Wbeam}
\end{equation}
where $P_{\ell}(\cos\theta)$ are the Legendre polynomial of index $\ell$ for the angular scale $\theta$, and $\textrm{PSF}(\theta,E)$ is the PSF as a function of angular distance $\theta$ and energy $E$. In the case of \fermilat, the $\textrm{PSF}(\theta)$ can be obtained for specific $E_{\gamma}$ values with {\tt gtpsf} tools, then the bin-averaged beam window function, $W^{beam}_E(\ell)$, can be obtained averaging the $W^{beam}(E,\ell)$ over the energy range considered, weighted by the UGRB intensity spectrum, which is approximately a power law with index -2.3 \citep{2015IGRB}:

\begin{equation}
W^{\gamma, beam}_E(\ell) = \frac{\int_{E_{min}}^{E_{max}} W^{\gamma, beam}(E,\ell) \frac{dN}{dE} dE}{\int_{E_{min}}^{E_{max}} \frac{dN}{dE} dE}~.
\end{equation}
\noindent
$W_{pix}$ is obtained with the HEALPix routine {\tt pixwin}, and is the same for both the \fermi and IceCube maps (since the pixeling order is the same), and it is shown as a gray dashed line in the left plot of Fig.~\ref{fig:Wbeams} in the Appendix.

The CAPS is binned in multipole in order to reduce/eliminate the correlation between adjacent multipoles in the $C_{\ell}$ spectrum (an effect induced by the presence of the mask). Following the procedure implied in \cite{UGRBaniso2018} and \cite{Fornasa2017}\footnote{The unweighted averaging procedure has been validated with Monte Carlo simulations by \cite{Fornasa2017} (see Section IV-A of that paper), and also applied in other similar cross-correlation analysis \citep[e.g.][]{Cuoco:2017bpv}}, the $\overline{C_{\ell}}$ value in the $\Delta\ell$ bin is computed as the arithmetic mean of the corresponding $C_{\ell}$ values. The errors of the binned CAPS is obtained from the associated covariance matrices following the procedure adopted by \cite{Fornasa2017}.

\begin{equation}
\Delta C_\ell = \sqrt{\sum_{\ell\ell'}\overline{V}_{\ell\ell'}/\Delta\ell^2}
\end{equation}
where $\overline{V}$ is the covariance matrix given by \textit{PolSpice} corrected by the window function: $\overline{V}_{\ell\ell'} = V_{\ell\ell'} W^{-2}_\ell W^{-2}_{\ell'}$. The covariance matrix computed with \textit{PolSpice} ignores effects due to non-Gaussian contributions. In this work, the cross-correlation measurement is dominated by the shot-noise terms and we expect higher order effects from non-Gaussian contributions to be negligible \citep{2015MNRAS.448.2854C}.

As demonstrated in \cite{UGRBaniso2018} the UGRB angular power spectrum is described by a constant function across the energy range considered in this work. According to a simple interpretation inspired by the halo model \citep{2002PhR...372....1C}\footnote{In this scenario the cross-correlation signal is attributed to the sum of a compact 1-halo term (tracing the intra-halo correlation), which is constant as a function of the multipoles, and a more extended 2-halo term, a decreasing function of the multipoles (tracing the inter-halos cross-correlation and representing the signature of correlation with the LSS).}, such observations can be interpreted with a dominant 1-halo term component produced by the emission of isotropically distributed point-like sources. Among other possible contributors, blazars are the absolute dominant population producing the observed anisotropy in the UGRB \citep{UGRBanisoInterpr}. This supports the choice to model the CAPS as a constant 1-halo term with the goal to assess the contribution of unresolved blazars to produce a cross-correlation signal.
The cross-correlation 1-halo term, usually denoted as \cp, is computed by fitting the measured CAPS with a constant by minimizing the $\chi^2$ function defined as 
\begin{equation}
    \chi^2 = \Delta^T~\overline{V}_{\ell\ell'}^{-1}~\Delta~~~{\rm where }~~~\Delta^T = (C_{\Delta\ell,1} - \rm{C_P}, ..., C_{\Delta\ell,N} - \rm{C_P})~.
\end{equation}
\noindent
In the fit, we do not consider multipoles below 20 to further exclude any possible contamination from large-scale residuals in the \fermilat{} maps due to mis-modeling of the Galactic foreground and/or the \fermilat{} exposure uncertainty. At high multipoles we are limited by the IceCube PSF, which we estimate to be 0.47 degrees at 68$\%$ containment angle for high-energy events. This limits the maximum multipole we can consider in this study to $\ell\approx 380$. More details on the estimation of the IceCube PSF profile is given in Appendix~\ref{app:A}.

\subsection{Data Selection}
The UGRB anisotropy energy spectrum as measured by \cite{UGRBaniso2018} shows a highly significant ($>4\sigma$) detection of anisotropies from point-like sources in the energy range between 1 and $\sim$25 GeV. In this work we only focus on this energy range for the gamma-ray data selection, and we work in four different energy bins (1-2, 2-5, 5-10, and 10-25 GeV). This energy-resolved study is allowed by the large photon statistics of the LAT data. As in \cite{UGRBaniso2018}, we consider a sub-selection of events (and corresponding response functions) with better angular resolution. This selection corresponds to {\tt SOURCEVETO} event class and {\tt PSF1+PSF2+PSF3} event type. In this study we use 12 years of \fermilat{} Pass 8 data. 

We use IceCube's 10~year public data release~\citep{IC10yr} derived from a recent time-integrated point source search~\citep{10yrpvalue}. This data release, taken between 2008 and 2018, includes a complete set of instrument response functions (``IRFs'') describing the reconstruction behavior of the detector as a probability mapping $P\left(E_{proxy}, \Psi, \sigma \mid| E_\nu, \delta \right)$ where $E_{proxy}$ is the reconstructed proxy for the energy, $\Psi$ is the directional reconstruction error, $\sigma$ is the estimated directional uncertainty, $E_\nu$ is the true neutrino energy, and $\delta$ is the source declination\footnote{Details on the meaning of each parameter are provided on IceCube's data release page and in a README file included in the release itself.}. Effective areas binned in true neutrino energy and declination are provided for several ``detector seasons'' corresponding to different versions of the detector or processing chain (``IC40'', ``IC59'',``IC79'', ``IC86-I'' and ``IC86-II+''). Instrument response functions allow users to map neutrino energy and source declination to reconstructed energy proxy, angular uncertainty, and point spread function (PSF) for each season. Reconstructed energy proxies, directions, and angular uncertainties for each observed event are also provided in the data release. To simplify calculations, we consider only the final detector season, ``IC86-II+'', spanning six years of data with uniform effective area and IRFs.

\subsection{Real data maps}\label{sec:real}
The \fermilat{} UGRB maps have been obtained following the same procedure as in \cite{UGRBaniso2018}, namely finely binning the data in energy (32 \textit{micro} logarithmic bins between 1-25 GeV) to produce intensity maps and then sum the final maps into \textit{macro} bins; as mentioned in Section~\ref{sec:wherehow}, we bin the data in four logarithmic energy bins between 1 and 25 GeV. We used version v10r0p5 of the \textit{Fermi Science Tools} to generate all-sky intensity maps in HEALPix format (order 8) as described in Section II of \cite{UGRBaniso2018}. As an example, Fig.~\ref{fig:masks} (middle panel) shows the \fermilat{} UGRB intensity map in the energy bin 2-5 GeV, where the gray areas are masked away as described in Section \ref{subsec:mask}. 

The procedure to subtract the residual Galactic foreground emission outside the masked region is detailed Section I of the Supplemental Online Material of \cite{UGRBaniso2018}. In this work we use the Galactic diffuse emission model {\tt gll\_iem\_v7.fits}\footnote{\url{https://fermi.gsfc.nasa.gov/ssc/data/analysis/software/aux/4fgl/Galactic_Diffuse_Emission_Model_for_the_4FGL_Catalog_Analysis.pdf}}.
% to the data maps (considering only unmasked pixels); the best-fit normalization was found to be around 0.95 but compatible with 1 within 1$\sigma$ and is then applied to the model template and then subtracted to the data maps. Note that this procedure is performed over the \textit{micro} energy bins. 
For the auto-correlation analysis in \cite{UGRBaniso2018}, the authors test that any residual contamination from the Galactic foreground is negligible above multipole 50: the angular power spectrum flattens after the foreground subtraction in the range of multipoles of interest (see their Fig. 3, left panel, of the Supplemental material). We repeated the test for our energy bins, confirming that this is still the case. In our cross-correlation study, therefore, we do not expect any significant residual contamination from the Galactic foreground.\\

The IceCube data map are obtained by filling an order 8 HEALPix map with the IceCube data provided in \cite{IC10yr}, selecting data taken in the 6 years between 2012 and 2018. In order to account for the widely varying IceCube response in declination, we choose to use a fluctuation map $\Xi$ for the neutrino data. We define the neutrino fluctuation map by normalizing the trial map for each HEALPix band in declination

\begin{equation}
\label{eq:ICfluct}
    \Xi_{pix}^{\delta} = \frac{N^\delta_{pix} - \left<N\right>^{\delta}}{\left<N\right>^\delta}
\end{equation}
\noindent
where $N^{\delta}_{pix}$ is the number of events observed in HEALPix pixel $pix$ at a declination of $\delta$ and $\left<N\right>^\delta$ is the average count over that declination. In Fig.~\ref{fig:masks}, the bottom row maps illustrate an example of IceCube simulated counts map (on the left) and the derived fluctuation map (on the right).
Fig.~\ref{fig:masks}, right panel, illustrates the IceCube fluctuation map with the mask applied (gray areas).

\subsection{Masking}
\label{subsec:mask}
The IceCube effective area for neutrino-like events has a strong dependence on the declination \citep{IC10yr}. In the Southern hemisphere ($\delta<-5^\circ$), the IceCube sky is dominated by muons produced in atmospheric air showers. Cuts are applied to remove these backgrounds from the Southern sky, resulting in a high energy threshold. In the Northern hemisphere ($\delta\geq-5^\circ$), Earth blocks atmospheric muon events from reaching IceCube detectors, allowing a lower energy threshold to be used. Instead of muons, atmospheric neutrinos generated in air showers provide an irreducible background in IceCube's Northern sky. Because the neutrino cross-section increases with energy Earth limits the number of high energy neutrino events visible in the Northern sky. 

In order to optimize our sensitivity to correlations between IceCube's neutrinos and the UGRB, we first study IceCube's expected response as a function of neutrino energy and arrival declination. We expect the UGRB-correlated neutrinos to contribute to IceCube's unresolved astrophysical diffuse flux, so we first compute the expected astrophysical neutrino events. To this end, we weight each energy and declination bin of IceCube's effective area using simple power law parametrized as in IceCube's most recent fit~\citep{Astronuflux},

\begin{equation}
\frac{d\Phi_\nu}{dE_\nu}\left(E_\nu\right)=1.44\times10^{-18} \left(\frac{E_\nu}{100~{\rm TeV}}\right)^{-2.37}~{\rm GeV}^{-1}{\rm cm}^{-2}{\rm s}^{-1}{\rm sr}^{-1}.    
% \Phi^{\nu_\mu+\bar{\nu}_\mu}
\end{equation}

We use these expected event counts to sample from the provided IRFs, producing $10^4$ realizations of the IceCube astrophysical diffuse flux binned in reconstructed energy proxy and direction assuming a uniform distribution across each IRF bin. These sampled events are compared to the observed events, building a map of the expected astrophysical contributions to each bin shown in Fig.~\ref{fig:ICexpected}. We see a strong divide between the expected astrophysical purity of the Northern and Southern sky, with the Northern sky reaching a purity of 10\% or higher at high energy proxies. In contrast, the Southern sky purity rarely breaches 1\%. 

We conclude that the Northern sky provides a significant advantage for astrophysical searches. We therefore limit our search to $\delta>-2^\circ$. Note that this is more stringent than the IceCube definition of the northern hemisphere to prevent from edge effects due to the sudden drop of the instrument effective area around declination $-5^\circ$.

We use scrambled neutrino data in background generation to match IceCube's standard methods \citep{10yrpvalue}. This procedure breaks near the poles due to limited event statistics, so we additionally mask out the polar cap ($\delta>-75^\circ$). 

\begin{figure}
    \centering
    \includegraphics[scale=0.4]{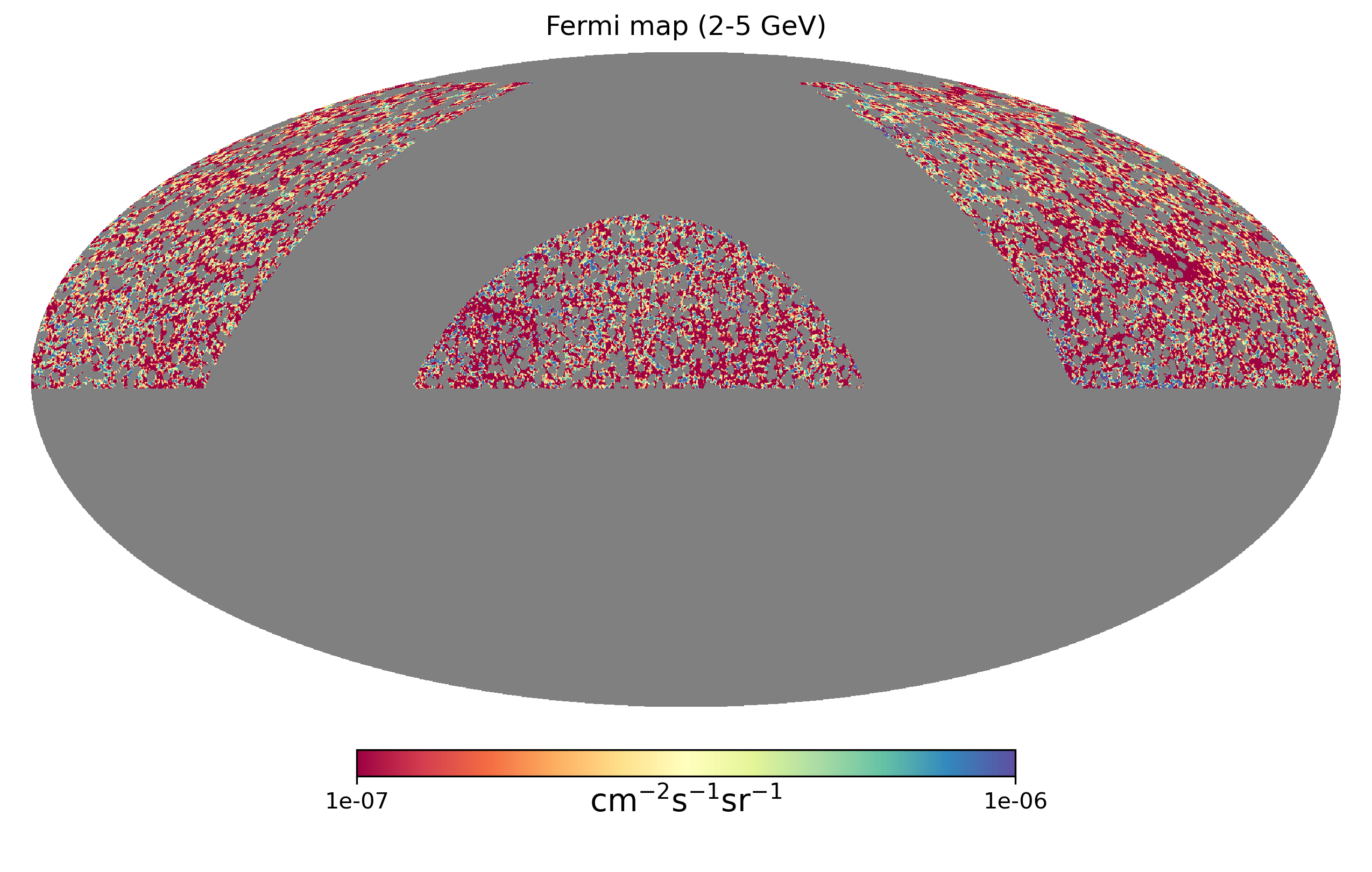}
    \includegraphics[scale=0.4]{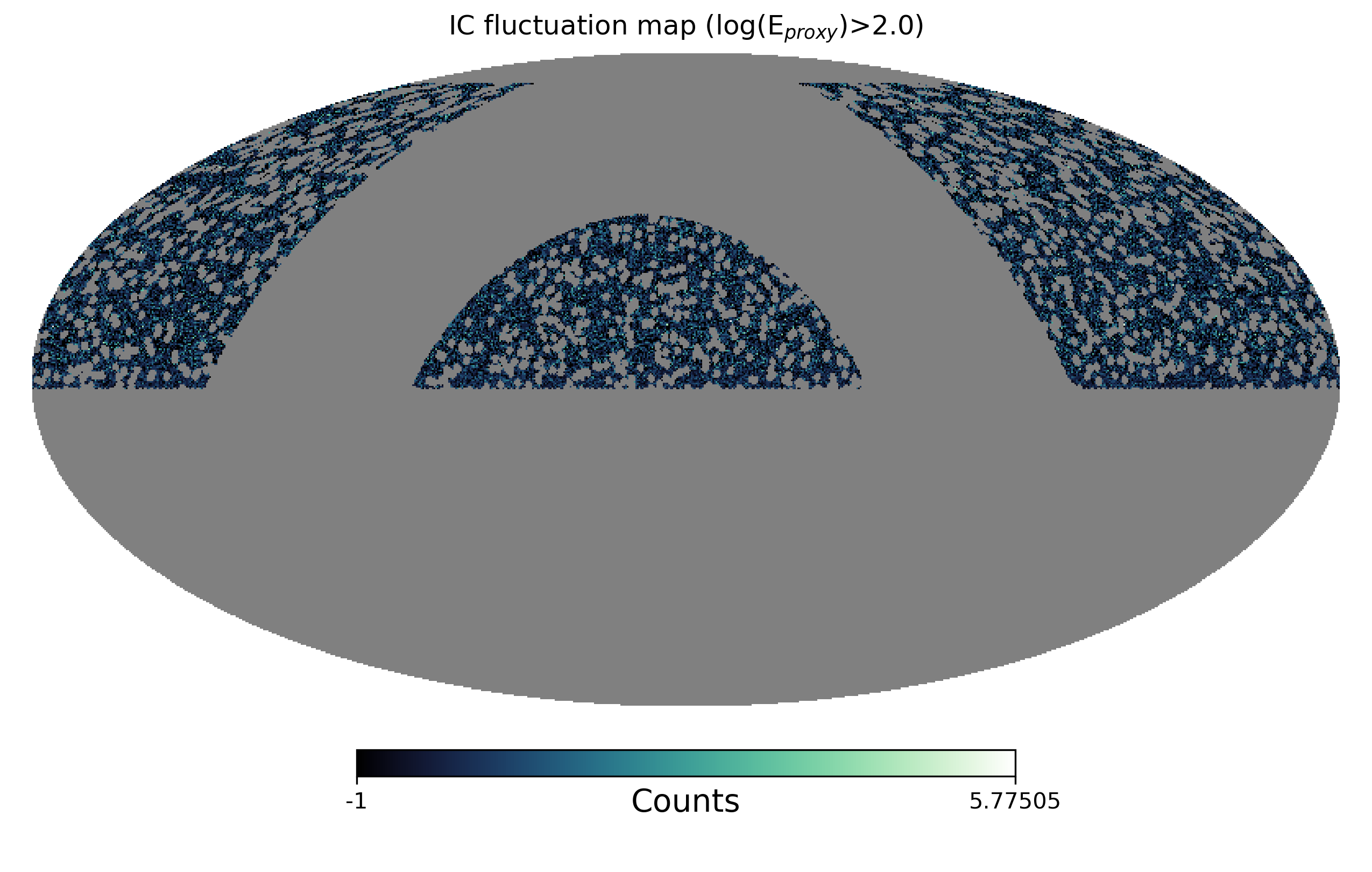}
    \includegraphics[scale=0.4]{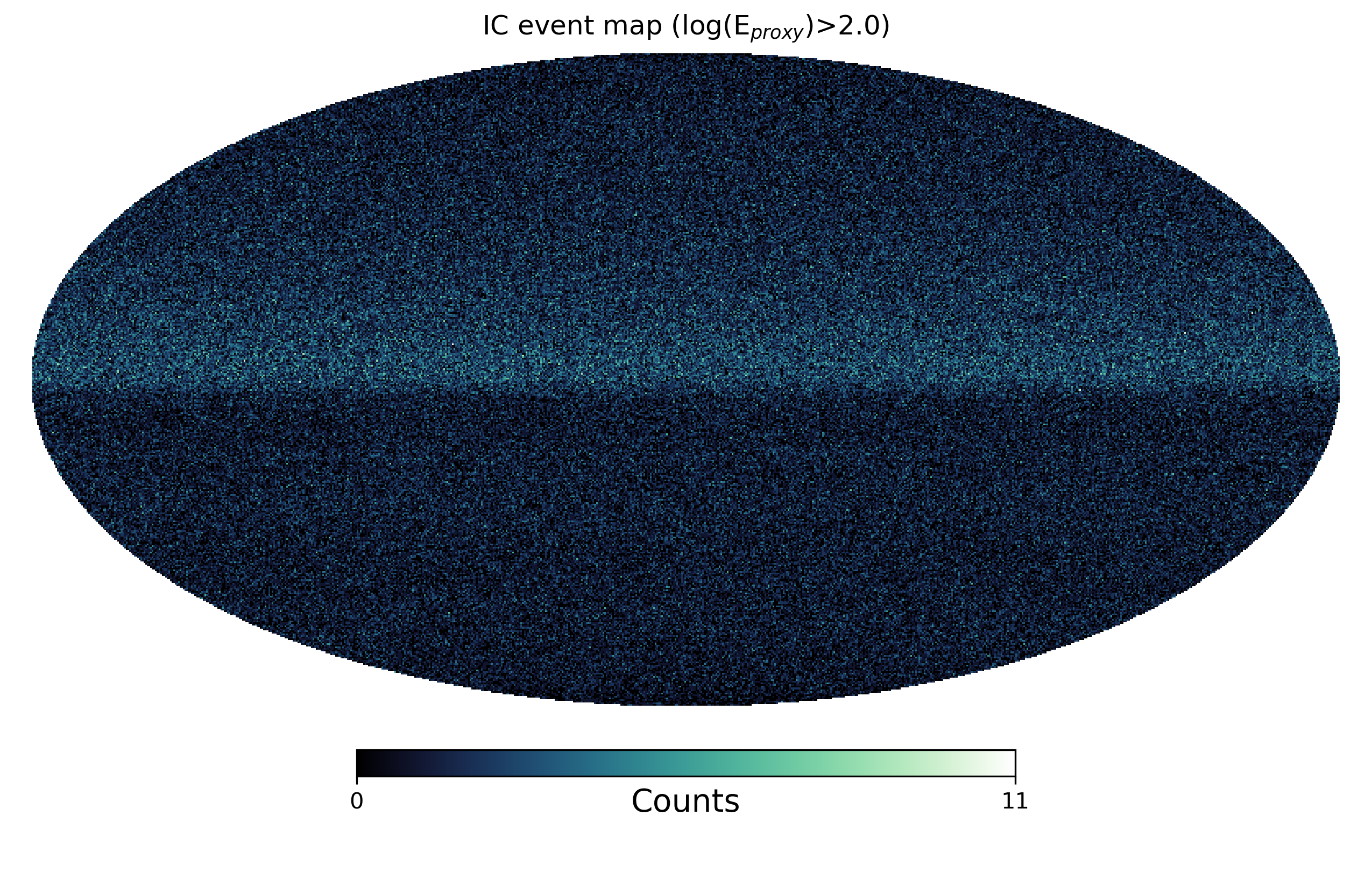}
    \includegraphics[scale=0.4]{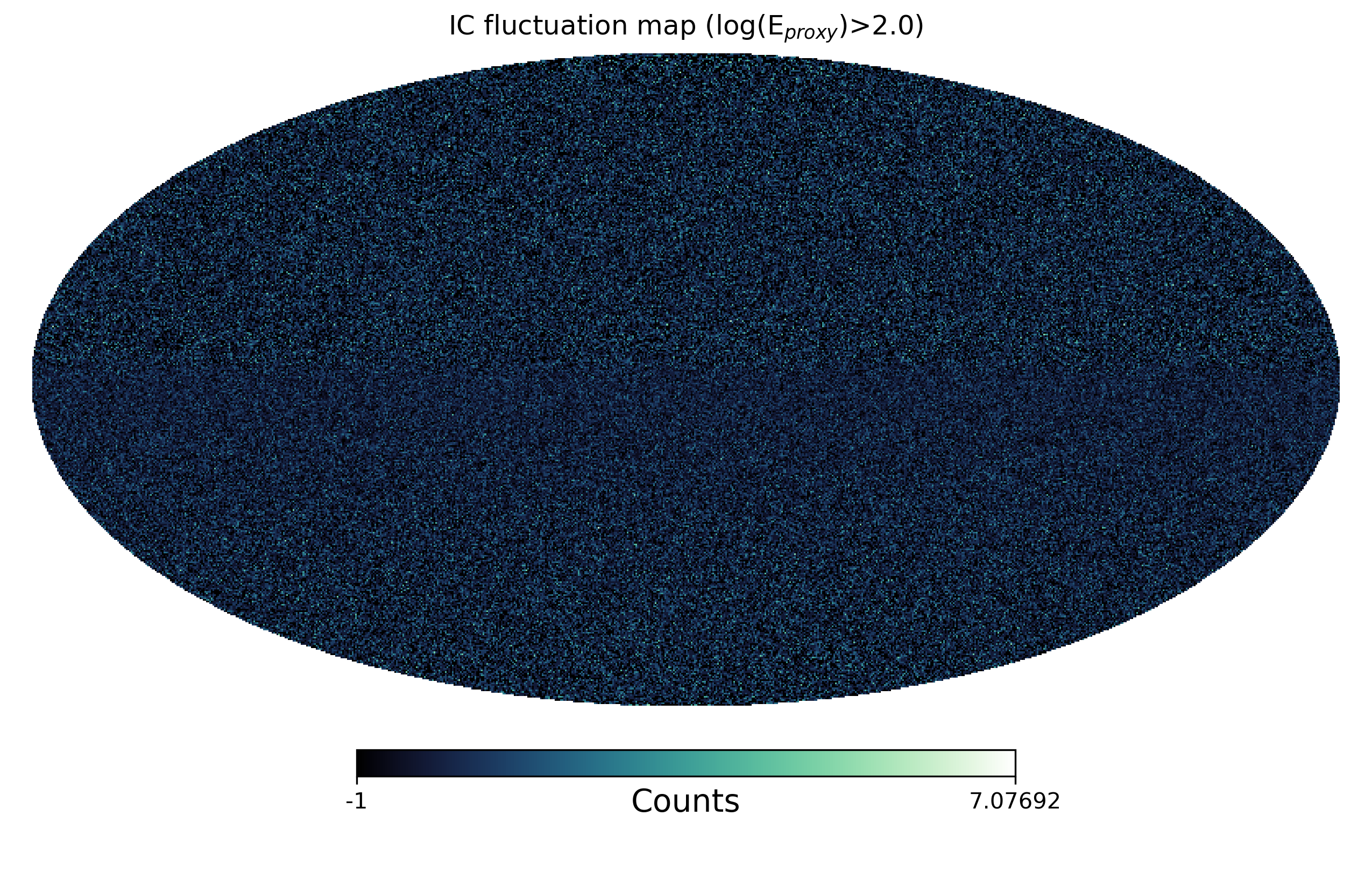}
    \caption{Top-Left: Example of \fermi map in the 2-5 GeV energy bin. Top-Right: IceCube (IC) counts fluctuation. For both maps in the top panle the gray areas show the mask applied, which is the combination of the southern hemisphere mask and the Galactic plane plus 4FGL sources mask. Bottom-Left: Example of simulated IceCube counts map. Bottom-Right: Fluctuation IceCube event map derived from the simulated counts map shown on the left. All the maps reported here are in celestial coordinates.}
    \label{fig:masks}
\end{figure}

\begin{figure}
    \centering
    \includegraphics[width=11.5cm, height=5cm]{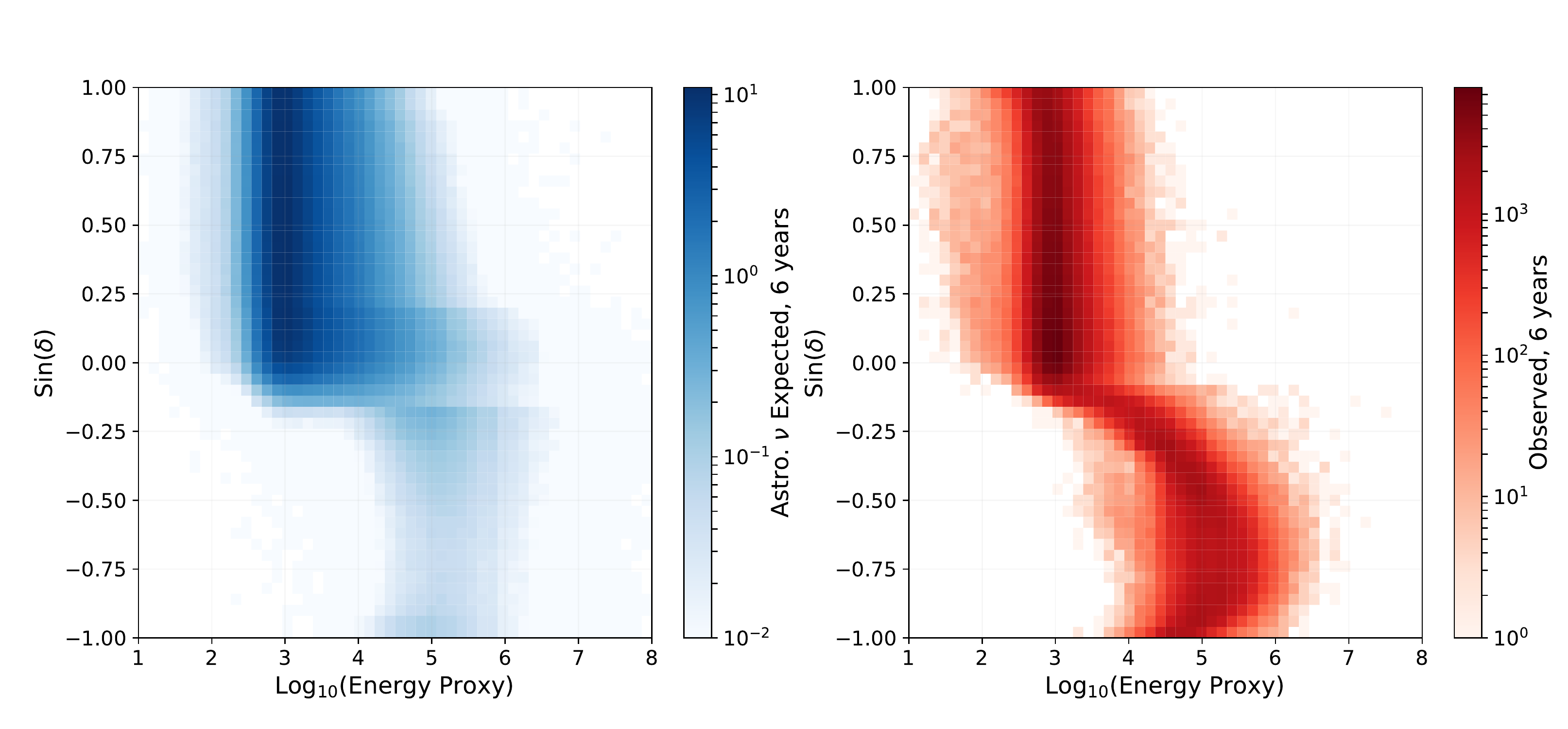}
    \includegraphics[width=5cm, height=4.85cm]{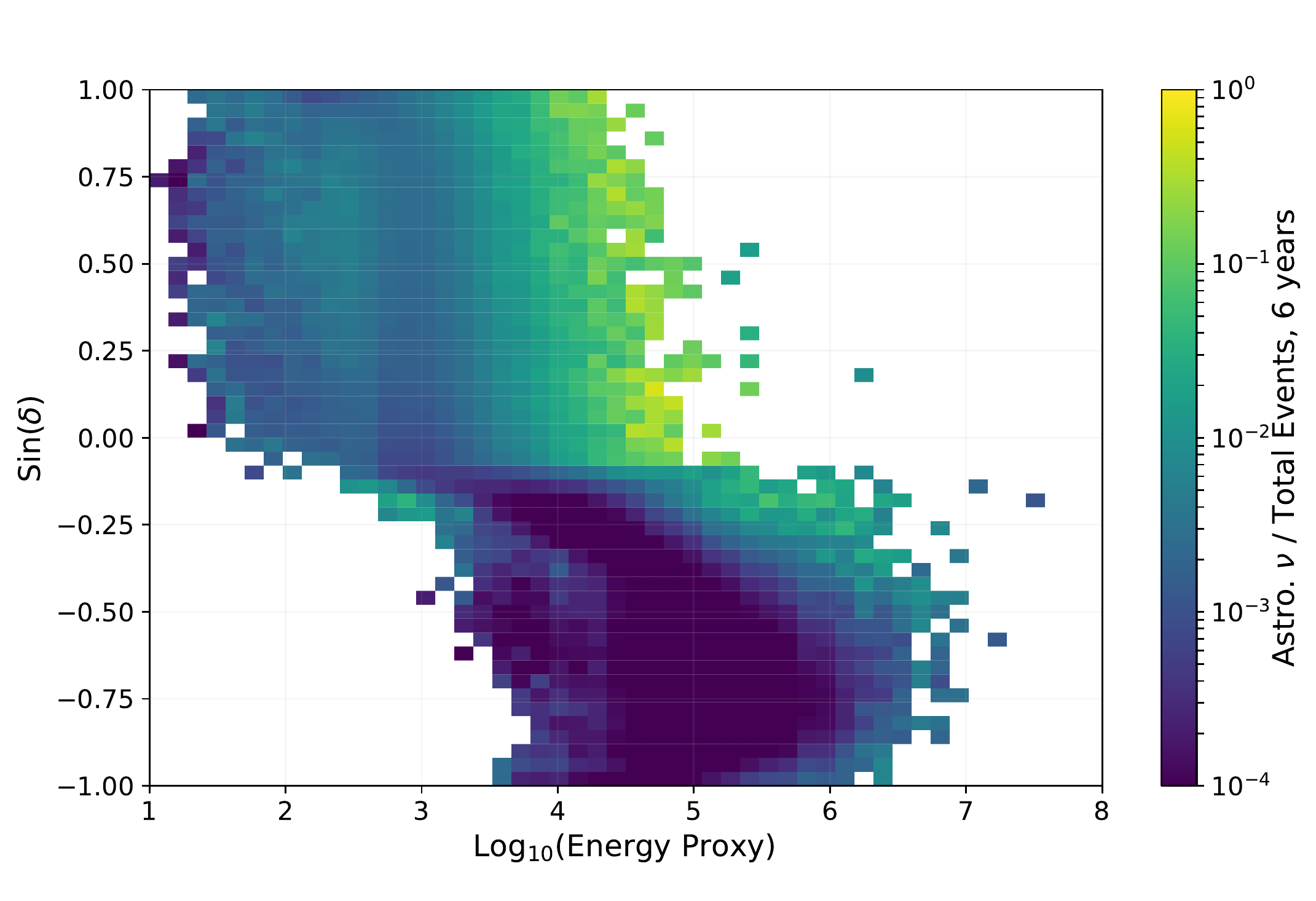}
    \caption{Quick study of the IceCube sensitivity to astrophysical neutrino signal considering the public data between 2012 and 2018. Left: Expected astrophysical neutrino events obtained assuming the latest astrophysical flux measurement by \cite{IClatestastro}. Middle: observed neutrino events as provided by the latest IceCube data release \citep{IC10yr}. Right: Ratio between the expected astrophysical neutrino events and the total expected ones: the Norther hemisphere is better suited for neutrino signal searches. In all plots Energy Proxy is expressed in GeV.}
    \label{fig:ICexpected}
\end{figure}

Mimicking the procedure in \cite{UGRBaniso2018}, we apply a 1 degree radius disk-like mask around the resolved gamma-ray sources listed in the 4FGL-DR3 \citep{4FGLDR3} and we mask the sky region within 25 degrees from the Galactic plane. The resulting total mask leaves free about 20\% of the sky and is shown in Fig.~\ref{fig:masks} (gray regions). Masking 25 degrees around the galactic plane is a conservative choice also adopted in several other cross-correlation studies \citep{Cuoco:2017bpv, AmmazzalorsoDES, Ammazzalorso:2018evf}. As discussed later in Section~\ref{sec:sim_maps}, mismodeling of the foreground emission to subtract from the data could result in some contamination. Conservatively masking this emission has demonstrated to be a good procedure to assure subdominant contamination from background subtraction. Furthermore, removing the majority of the bright Galactic emission along the Galactic plane considerably reduces the noise, which affects the variance of the cross-correlation measurement.

The southern sky, north polar cap, Galactic plane and 4FGL-DR3 masks are combined into a single mask. All maps are generated in HEALPix format, and we use the \textit{healpy} python package to handle and analyze them. Because IceCube's angular uncertainty is of O($0.5^\circ$) \citep{IC10yr}, we choose to work with a relatively coarse pixelization defined by an NSIDE of 256 (order 8) corresponding to an average pixel resolution of approximately 0.2 degrees. 

\section{Simulations}
\label{sec:sim}

This analysis searches for a cross-correlation signal between the gamma-ray intensity field (as seen by the LAT) generated by an unresolved population of blazars and a neutrino count fluctuation field (as seen by IceCube) from the same population of blazars assuming a neutrino production from \textit{p-p} interaction or \textit{p-}$\gamma$ interactions.
%The instrument response functions, the level of white noise in the data, and the statistics involved in this kind of analyses represent the main sources of uncertainty to this measurement.
Because we are correlating disparate measurements, the resulting value of \cp{} is difficult to interpret a priori.
We use simulations with known levels of correlation to both determine the sensitivity of our analyses to the presence of a signal and to convert our measurement of \cp{} into an upper limit on the fraction of the UGRB $\gamma$-ray flux produced in hadronic interactions.
In this Section, we describe the procedure used to build simulated \fermilat{} intensity maps and IceCube counts maps from a synthetic blazar population.

\subsection{Unresolved blazar population}
Our goal is to simulate a population of unresolved blazars that on one hand reproduces the expected $\sim$ 30\% of the UGRB intensity spectrum \citep{2015IGRB}, and on the other hand matches the measured UGRB anisotropy energy spectrum \citep{UGRBaniso2018}. To this end, the work from \cite{LeaSimBlazar} provides us with two important tools: (1) a simulation of the intrinsic blazar population (detected and undetected sources) that we can use to simulate both \fermi and IceCube maps; (2) a catalog of blazars detected from the simulations via a reliable detection pipeline optimized to recover the preliminary 4FGL catalog (the FL8Y), which we use to construct a mask for the simulated extragalactic sky (see next Section~\ref{sec:sim_maps}).  
The simulation campaign presented in \citet{LeaSimBlazar} was aimed at reproducing the observed spectral characteristics and statistics of the resolved extragalactic gamma-ray sources (i.e.,~blazars). Under the assumption that these sources are uniformly distributed in the $\gamma$-ray sky, they built blazar populations with: (i) flux distribution extending an order of magnitude below the \fermilat{} source detection sensitivity; (ii) an intrinsic source count distribution ({\it logN-logS}); and (iii)
%the authors performed an end-to-end Monte Carlo simulation, with the aim of reproducing the observed statistic of resolved extragalactic gamma-ray sources. They simulated a blazar population with intrinsic characteristics modelled as in \cite{Ajello:2015mfa}, namely using a double power law as spectral shape (eq. 1 of \cite{LeaSimBlazar}) and a Pure Density Evolution (PDE) model, which best describes the observed real source catalog. 
a double broken power-law intrinsic energy spectrum ($\frac{dN}{dE}$) for each source of the form:

\begin{equation}
\label{eq:simspec}
    \frac{d\phi_\gamma}{dE} = K \left[ \left( \frac{E}{E_b(\Gamma)}\right)^{\delta_1} + \left( \frac{E}{E_b(\Gamma)}\right)^{\delta_2}\right]^{-1},
\end{equation}

\noindent
with log($E_b(\Gamma)) = 9.25 - 4.11\Gamma$ being the energy in GeV of the spectral break and $\Gamma$ the power-law photon index of blazar's $\gamma$-ray spectrum as measured by the LAT \citep{Ajello:2015mfa}; 
% ($\Gamma$, extracted randomly from a Gaussian centered at 2.45); 
$\delta_1 = 1.7$ and $\delta_2 = 2.8$ are the spectral indices before and after the break energy, whose values have been found to reproduce the source-count distribution of the Third Catalog of Hard LAT Sources (3FHL, \cite{3FHL}). We refer the interested reader to \cite{LeaSimBlazar} for further details on the simulation of the blazar population. In this work, we consider the simulation built on the {\it logN-logS} modeled as a double broken power law (model 2 in Table 2 of \citealp{LeaSimBlazar}).

\begin{figure}[t]
    \centering
    \includegraphics[height=6cm,width=10cm]{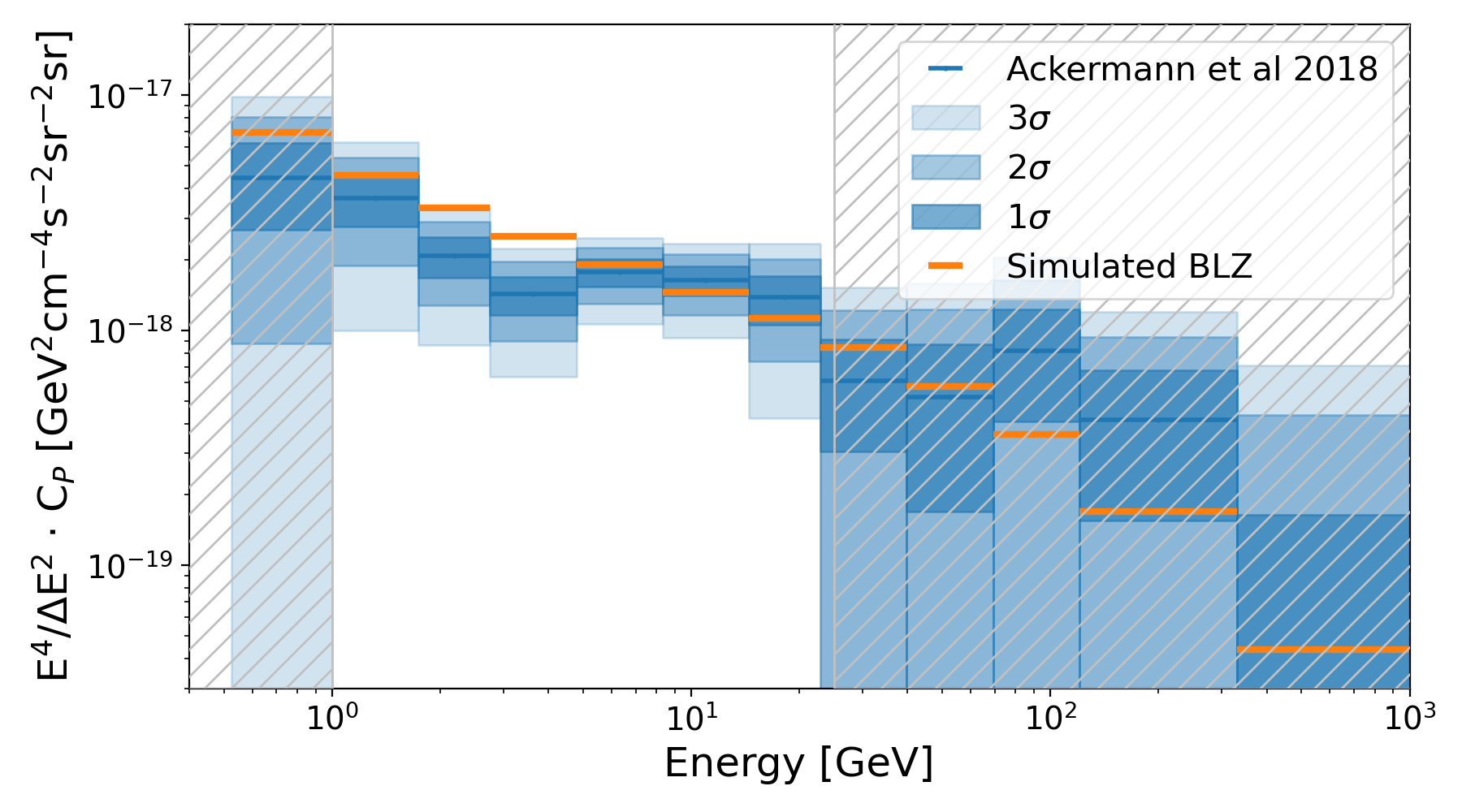}~~
    \includegraphics[height=6cm,width=7cm]{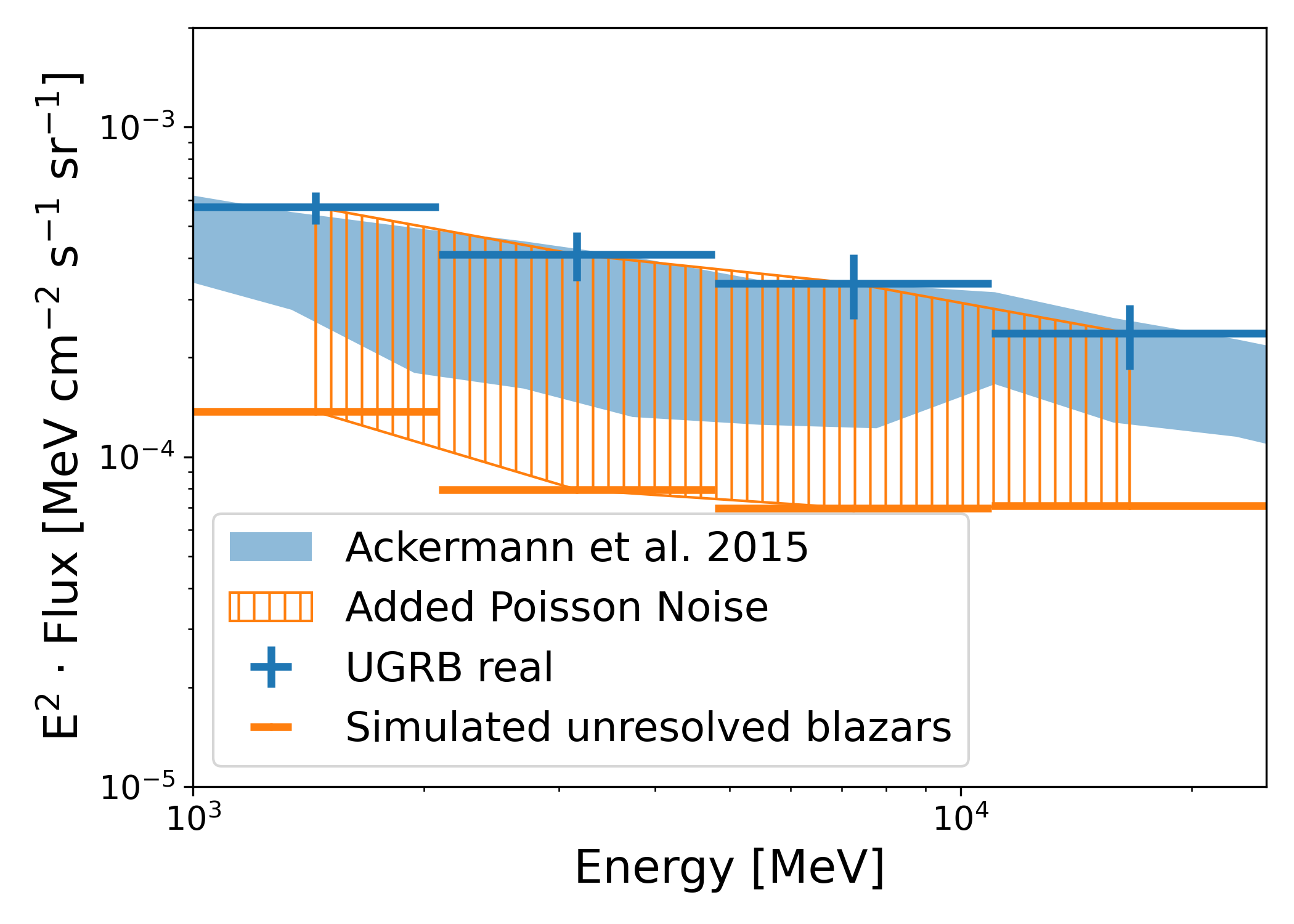}
    \caption{Left: Anisotropy power carried by the simulated unresolved blazar population (orange) compared to the measured UGRB anisotropy energy spectrum in \cite{UGRBaniso2018}. Right: Photon intensity of the simulated unresolved blazar population (orange lines), the photon intensity of the UGRBas measured from the real data maps (see Section~\ref{sec:real}) The orange dashed region marks the fraction of the intensity added to the simulated \fermi maps as poisson white noise in order to match the real data intensity in each energy bin. The blue shaded part is the systematic band of the UGRB energy spectrum as presented in \cite{UGRBaniso2018}. Note that all the intensities are estimated from the \textit{unmasked} region of the sky, according to the mask described in Section~\ref{sec:wherehow}}.
    \label{fig:Simtest}
\end{figure}

While \cite{LeaSimBlazar} extensively demonstrated that the population of detected simulated sources give an adequate representation of the real \fermilat{} extragalactic gamma-ray source population, we still need to make sure that also the unresolved regime is statistically representative of the real UGRB. We want to verify that the simulated unresolved blazar population brings an anisotropy power that matches the observed UGRB anisotropy energy spectrum. In order to do so, we first define a sky mask to cover the detected sources (covered with a disk of 1 degree radius) and the Galactic plane (25 degree bands above an below). Then we compute the cumulative anisotropy level for each energy bin, ${\rm C_P}(\Delta E)$, from all the sources that fall outside the masked region as

\begin{equation}
    {\rm C_P}(\Delta E) = \frac{1}{4 \pi f_{sky}}\sum_{{\rm src}}[\Phi(\Delta E)]^2
\end{equation}

\noindent
where $f_{sky}$ is the fraction of sky that is unmasked, $\Phi(\Delta E)$ is the integrated flux in the energy bin considered, and the sum runs over the unmasked sources. Fig.~\ref{fig:Simtest} (left panel) shows the level of anisotropy of the simulated unresolved blazars compared to the measured anisotropy energy spectrum by \cite{UGRBaniso2018}. The agreement is good and validates the use of the simulated blazar population by \cite{LeaSimBlazar} for our study. 

\subsection{Simulated IceCube neutrino maps}
\label{sec:simIC}
We generate signal neutrinos from the simulated list of resolved and unresolved blazars using the IceCube response functions provided in the latest release \citep{IC10yr}, and assuming that all blazars produce neutrinos in \textit{p-p}~interactions following the same spectra as the $\gamma$-rays. Such an assumption is rendered by extrapolating the $\gamma$-ray spectrum down to IceCube energies and converting the gamma-ray flux into a neutrino flux following the relationship between the all-flavor neutrino flux and the $\gamma$-ray flux from \cite{KeKota}:

\begin{equation}
\label{eq:p-p}
    E^2_\nu \frac{d\phi_\nu}{dE_\nu} \approx \frac{3}{2}\left(E^2_\gamma\frac{d\phi_\gamma}{dE_\gamma}\right) \Bigm| _{E_\nu\approx E_\gamma/2}
\end{equation}
\noindent
We divide the all-flavor neutrino flux by three ---implicitly assuming complete mixing of neutrino flavors--- since the IceCube release includes only contributions from muon neutrinos. The extrapolation of the gamma-ray spectrum implicitly assumes no energy cutoff or any additional harder component emerging at IceCube’s energies. In these latter cases, predicting the spectral shape at TeV energies is difficult due to the wide range of possible scenarios. As a result, we opted to extrapolate the power law defined at GeV energies. Variations in sensitivity produced by different power-law indices at IceCube energies are explored in Appendix \ref{app:B}.
%\textcolor{blue}{include a note about the broken power law being the same $\simE^-2.7$ for all sources at IceCube energies? MN: no, need: everything is explained in the Lea's paper}.
For each source $i$, we calculate the expected number of neutrino events, denoted by $\mu^{i}$, by combining the derived single-flavor neutrino fluxes, the IceCube livetime, $\Delta t$, and IceCube effective area, $A_{\rm eff}\left(E_\nu,\delta^i\right)$:
\begin{equation}
    \mu^{i} = \Delta t \; \int A_{{\rm eff}}(E_\nu, \delta^i) \left(\frac{d\phi_\nu}{dE_\nu}\right)^i\; \; dE_\nu
\end{equation}
\noindent
where $\Delta t\sim 6$ years for the ``IC86-II+'' seasons used here, and $\frac{d\phi_\nu}{dE_\nu}$ is derived using Eq.~\ref{eq:p-p} assuming the $\gamma$-ray spectrum in Eq.~\ref{eq:simspec}. The IceCube effective area is provided as averages $\bar{A}_{\rm eff}$ over discrete energy bins $[E_{\nu}^{j,min}, E_{\nu}^{j,max}]$, converting our calculation of $\mu^i$ into a summation
\begin{equation}
    \mu^{i} = \Delta t \; \sum_j \bar{A}_{\rm eff}^j\left(\delta^i\right) \int_{E_\nu^{j,min}}^{E_\nu^{j,max}} \left(\frac{d\phi_\nu}{dE_\nu}\right)^i dE_\nu
\end{equation}
The number of signal events added to each trial is drawn from a Poisson distribution assuming $\mu^i$ as mean value. Events are distributed around the source according to the point spread function embedded in IceCube IRF files. 
% , and the intrinsic neutrino energy for each event, $E^k_\nu$ is generated according to the power law extrapolated from the $\gamma$-ray intrinsic spectrum. 
%The signal neutrino events are produced for each effective area bin by Poisson-sampling a number of observed events from $N^{ij}_\nu$ assuming a mean of $\mu^{ij}_\nu(\Delta E_\nu^j)$. In the final map, for each source, the total number of simulated neutrino events is given by $N_\nu^i = \sum_j N^{ij}_\nu$. The reconstructed energy proxy and the estimated angular uncertainty are generated by choosing from the IceCube IRFs with probability $P\left(E_{proxy}, \sigma_{{\rm ang}} \mid| E^k_\nu, \delta^i \right)$, then sampled from a uniform distribution across the selected bin.

To produce simulated background measurements, we sample events from the data with Poisson fluctuations, then scramble the right ascension values of the sampled events. This procedure assumes that the backgrounds are local and azimuthally symmetric, that any potential signal would be defined by spatial coincidences which may be broken by scrambling, and that the total rate of events is dominated by atmospheric backgrounds. 

Signal events, if present, are added to the trial while removing an equal number of randomly selected background events, ensuring that the total number of observed events in each trial remains consistent with data. The final set of events in the trial are then binned in a HEALPix map.

The obtained counts map is converted in fluctuation map by following the same declination-dependent procedure adopted for the real data maps (Eq.~\ref{eq:ICfluct}).

\subsection{Simulated LAT UGRB map}\label{sec:sim_maps}

From the list of simulated unresolved sources, we generate simulated realizations of \fermilat{} UGRB maps in the four energy bins defined in the previous section. Each of these simulated maps is then cross-correlated with the trial simulated IceCube maps, allowing a study resolved in gamma-ray energy. This is advantageous because the final cross-correlation result may be affected by the signal-to-noise ratio of the gamma-ray maps, which improves with increasing energy, but also the intensity of the UGRB sources which decreases with increasing energy. Not knowing a priory which effect will dominate the measurement, binning in energy allows us to perform a blind sensitivity study.

%The intrinsic energy spectrum $\frac{dN}{dE}$ for each source, given by \cite{LeaSimBlazar} in units of cm$^{-2}$s$^{-1}$MeV$^{-1}$, is integrated between E$_{min}$ and E$_{max}$ defined by the energy bin edges, where 

%\begin{equation}
%    \frac{dN}{dE} = K \left[ \left( \frac{E}{E_b}\right)^{\delta_1} + %\left( \frac{E}{E_b}\right)^{\delta_2}\right]^{-1}
%\end{equation}

%$E_b = 9.25 - 4.11\Gamma$ being the break energy, $\Gamma$ the power-law photon index of blazar's $\gamma$-ray spectrum (relation from \cite{Ajello:2015mfa}), $\delta_1 = 1.7$ and $\delta_2 = 2.7$ which reproduce the source-count distribution of the Third Catalog of Hard LAT Sources (3FHL, \cite{3FHL}) as explained in \cite{LeaSimBlazar}.  

The integrated flux from each source, in units of ph~cm$^{-2}$s$^{-1}$ obtained by integrating Eq.~\ref{eq:simspec} from $E_{\rm min}$ to $E_{\rm max}$, is added to an initially empty HEALPix map in the pixel corresponding to the location of the simulated sources. The flux is converted to intensity in units of cm$^{-2}$s$^{-1}$sr$^{-1}$ dividing by the pixel area. 
%Source by source 
For every source, we apply a convolution with the \fermilat{} PSF, which is function of the energy and the separation angle $\theta$ form a given sky direction. In agreement with the data selection (Section~\ref{sec:real}), we consider the responses for {\tt SOURCEVETO} ({\tt PSF1+PSF2+PSF3}) event selection and we average over the energy bin weighting by the source spectrum.

Once all sources are added in the map, we add a white noise component to match the total UGRB intensity. This is achieved by injecting an isotropic component in the map with value equal to the difference between the true measured total intensity and the total intensity flux given by the simulated unresolved blazars. We illustrate the additional noise component with the hatched orange band in Fig.~\ref{fig:Simtest} (right panel). Notice how the total intensity from the simulated blazars represents roughly 30\% of the total UGRB flux, in agreement with the expectations. At this point we derive a counts map by multiplying by the \fermilat{} exposure map obtained while reducing the real data (see Section~\ref{sec:real}). A pixel-by-pixel Poisson randomization is applied to the counts map and then converted back to intensity by dividing by the same exposure map.

As discussed earlier, we subtract the Galactic foreground emission from the real Fermi-LAT data maps using a model template. Ideally, the subtraction is perfect, however, there might be some residuals and/or artifacts due to an imperfect foreground modeling, which may affect the anisotropies at different scales. This effect, which is particularly relevant for UGRB autocorrelation analysis, can be largely neglected in the case of cross-correlations.
%because of the very conservative mask we adopt to cover the Galactic plane emission. 
% For the energies considered, we tested that beyond multipole 50 such contamination is negligible even when computing an auto-correlation of the \fermilat{} maps. 
Furthermore, by injecting the Poisson noise component in the simulated maps that matches the total intensity of the real data maps (and hence including any possible small-scale artifact due to mis-modeling of the foreground), we ensure that the correct amount of variance is reproduced when computing the CAPS.

\section{Results}
\label{sec:res}
In this section we first describe the results we obtain from the simulation campaign. Then we unblind the analysis by computing the cross-correlation of the real data maps.

The number of trials of our simulation campaign, which aims to estimate the sensitivity of this analysis technique to detect a neutrino signal from gamma-ray unresolved blazars, is given by the number of realizations of simulated IceCube event map. We generated 10,000 IceCube map realizations for a range of variations in signal strength, $\kappa$

\begin{equation}
\label{eq:kappa}
    \left(\frac{d\phi_\nu}{dE_\nu}\right)^{inj} = \kappa \left(\frac{d\phi_\nu}{dE_\nu}\right)
\end{equation}

\begin{figure}
    \centering
    \includegraphics[width=\textwidth]{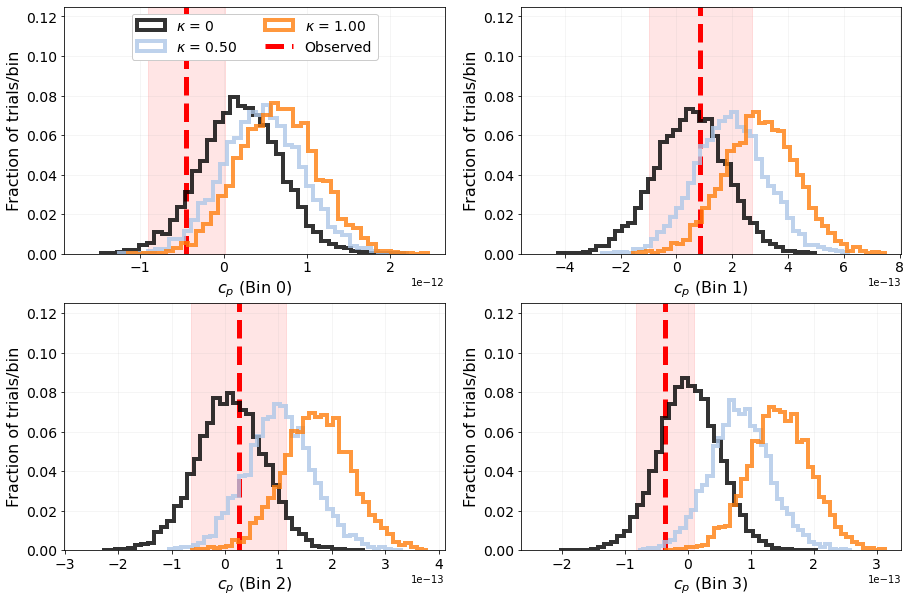}
    \caption{Trials histograms for the four \fermilat{} energy bins considered in this study. The black line corresponds to the null hypothesis of having only background events in the IceCube data (no correlation expected). The colored lines are the trials distributions for different values of $\kappa$. As we increase the amount of correlated signal, the value of ${\rm C_P}$ increases for each energy bin.}
    \label{fig:trialsdistrib}
\end{figure}

The simulated value of $\kappa$ ranges from 0 - corresponding to the null hypothesis of no correlated signal injected - to ten.
The obtained trial distributions for each signal strength are used as likelihood functions to derive the analysis sensitivity. The trials distributions of the \cp{} for each \fermilat{} energy bin considered are shown in Fig.~\ref{fig:trialsdistrib}.
%\textcolor{orange}{We haven't actually talked about doing the comparison to each fermi bin until now, have we?}. 
The distribution for the null hypothesis of having only background events is marked in black, while the colored distributions are the alternate hypotheses injecting different amount of signal by varying the parameter $\kappa$. As expected the ${\rm C_P}$ distributions are approximately Gaussian and therefore we define the likelihood functions as the best-fit Gaussian to the trial distribution for each injected flux:

\begin{equation}
    \mathcal{L}^{B}({\rm C_P}) = \frac{1}{\sqrt{2\pi}\sigma_B}{\rm e}^{\frac{({\rm C_P - C_P^B})}{2\sigma_B^2}} \quad \quad \quad
     \mathcal{L}^{S_\kappa}({\rm C_P}) = \frac{1}{\sqrt{2\pi}\sigma_\kappa}{\rm e}^{\frac{({\rm C_P - C_P^{S_\kappa}})}{2\sigma_\kappa^2}}
\end{equation}
where $C_P^B$ and $C_P^{S_\kappa}$ are the Gaussian mean value of the trial distributions for the null and alternate hypotheses, respectively.
We interpolate the means and variances as a function of $\kappa$ to obtain a continuous likelihood as a function of injected signal. A test statistic is calculated from the delta log-likelihood of the likelihood functions evaluated at the median value as:

\begin{equation}
    {\rm \Delta_\kappa} = - 2 \left[ \log\mathcal{L^B}({\rm C_P^B}) -  \log\mathcal{L^{{\rm S_\kappa}}}({\rm C_P^{S_\kappa}})\right]
\end{equation}

\begin{figure}[t]
    \centering
    \includegraphics[scale=0.35]{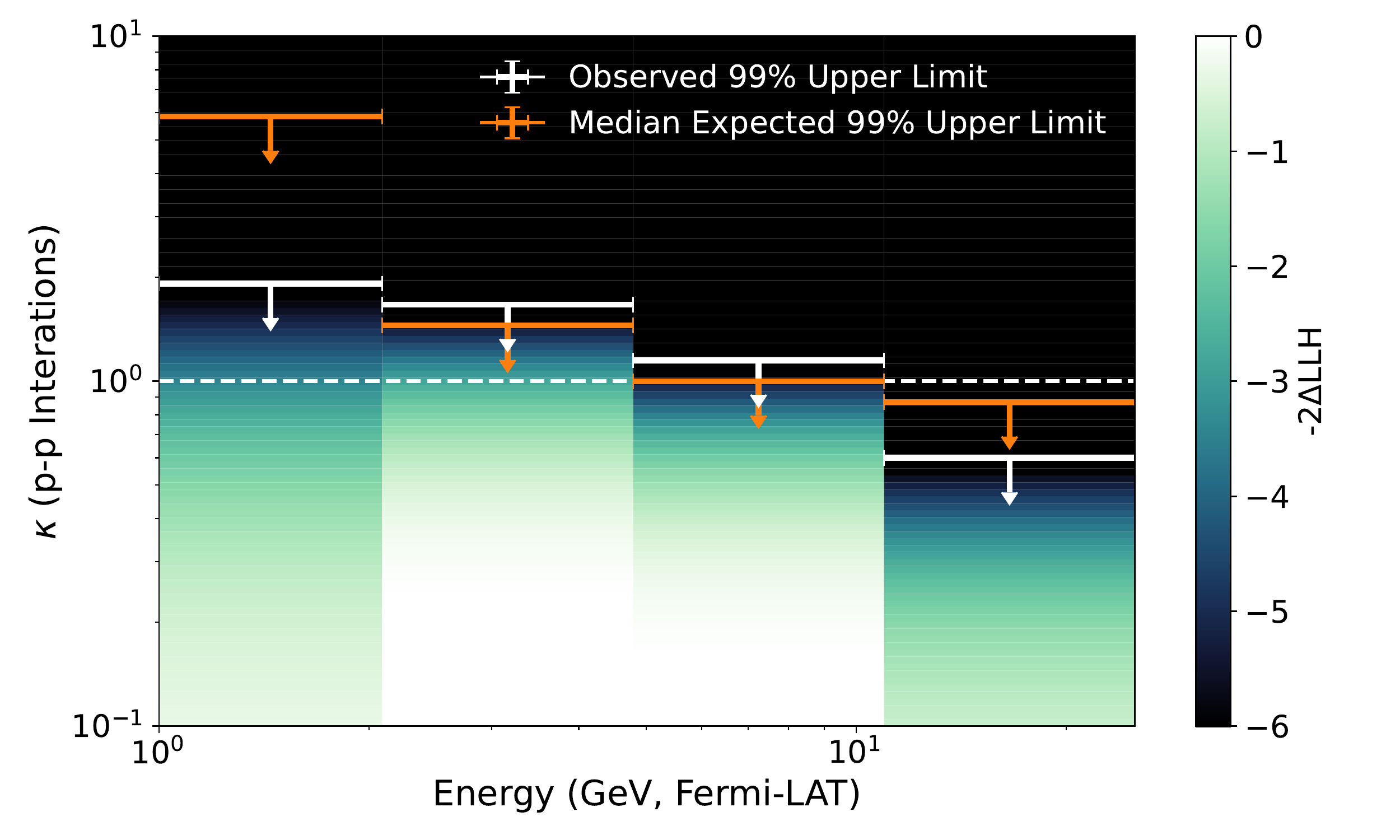}
    \includegraphics[scale=0.35]{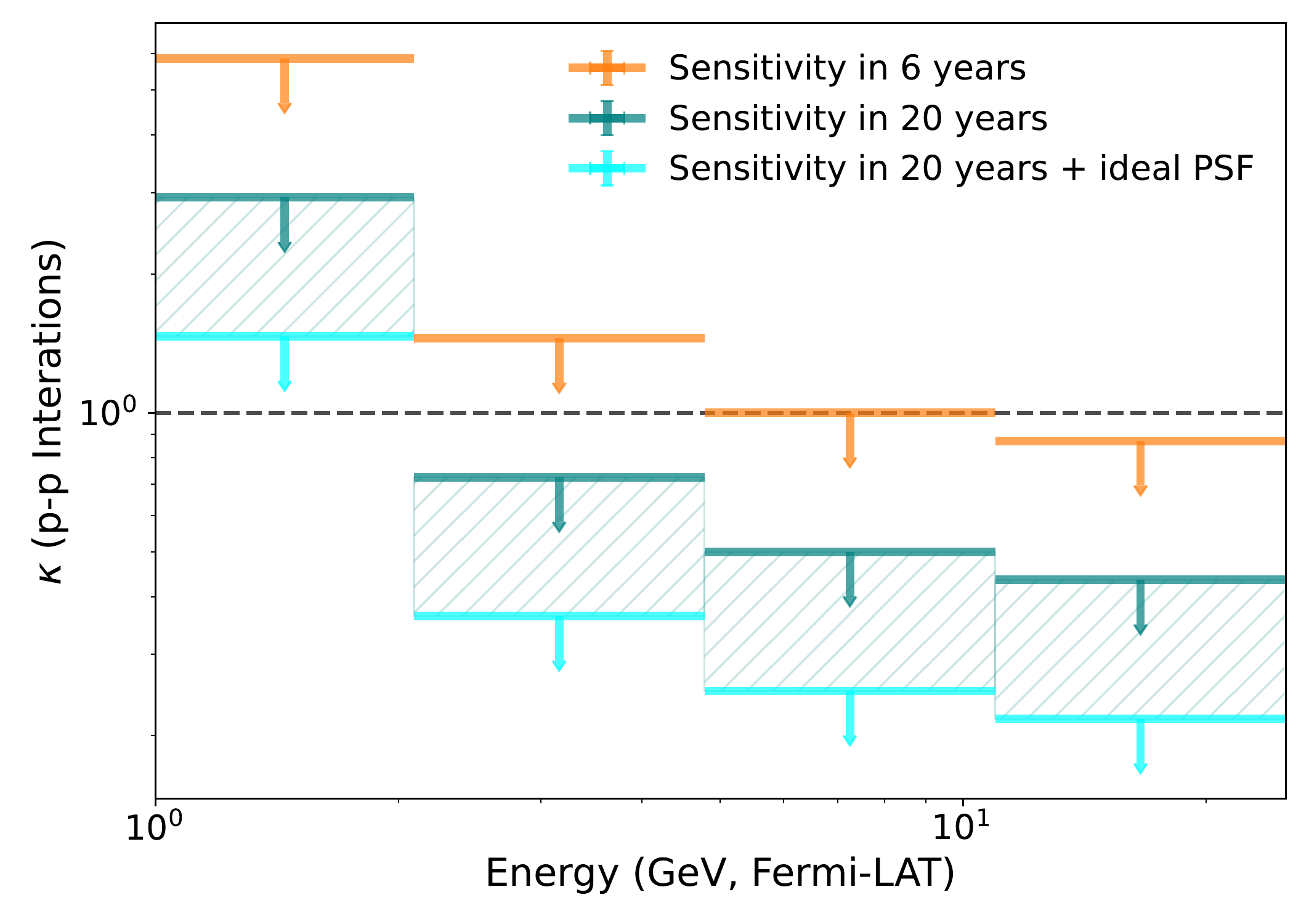}
    \caption{Left: Expected sensitivity to neutrino signal from \textit{p-p} interaction production channel and unblinded upper limits. In orange we mark the 99\% C.L. upper limit. In white we mark the data-driven 99\% C.L. upper limit. The dashed thin horizontal lines is added to mark the reference value $\kappa=1$. Right: Sensitivity projected to 20 years of IceCube data for the \textit{p-p} interaction. The \textit{p-}$\gamma$ neutrino production channel can be obtained by scaling the \textit{p-p} interaction sensitivities by a factor of 2.}
    \label{fig:sens}
\end{figure}

Noting that the background and signal models are nested and linear in the fitted parameter $\kappa$, our test statistics $\rm \Delta_\kappa$ is $\chi^2$ distributed with one degree of freedom \citep{Wilks:1938}. We derive the sensitivity at 99\% confidence level as the $\kappa$ value where the $\Delta$ = 6.67. The 99\% sensitivity is shown in Fig.~\ref{fig:sens} (orange arrows). Note how the highest energy bin is the most sensitive to a cross-correlation signal. This can be attributed to the higher signal to noise ratio in the \fermilat{} maps at higher energies (less white noise with respect to the blazars emission). 

% This sensitivity study tells us that we can expect to be sensitive to neutrino signal coming from p-p production mechanisms only when considering the highest energy bin, where we could recover a $\kappa=0.66$ at 99\% confidence level. For p-$\gamma$

We explored the hypothesis of neutrino produced via proton-$\gamma$ interactions. The procedure to generate the IceCube simulated maps is the same as the one described in Section~\ref{sec:sim} except that Eq.\ref{eq:p-p} now reads:
\begin{equation}
\label{eq:pgamma}
    E_\nu^2 \frac{dN_\nu}{dE_\nu} \sim \frac{3}{4}\left( E_\gamma^2 \frac{dN_\gamma}{dE_\gamma} \right)|_{E_\nu\sim E_\gamma/2} ~.
\end{equation}
\noindent
which is the all-flavor flux assuming perfect mixing of neutrinos after oscillation. Once again we divide it by 3, because we are considering only muon-neutrinos.
For completeness we also performed simulations considering only IceCube events with $\log(E_{{\rm proxy~}}[{\rm GeV}])>4$ in order to see whether enhancing the ratio between astrophysical neutrinos and atmospheric neutrinos could lead to a more promising study. The result, however, was a non-constraining sensitivity because the IceCube statistics in the northern hemisphere above those energies is too low to pick up any cross-correlation signal.\\ 

%90% Limits: [0.7926972567354073, 1.4512212125906352, 1.2621982022284755, 0.32754348943681133]
%90% Sensitivity: [1.748001088264063, 0.8304421333491658, 0.6281990195348748, 0.5215418623011345

The results of the cross-correlation with real maps are reported in Tab.~\ref{tab:realCp}. The ${\rm C_P^{{\rm real}}}$ values are all compatible with zero within the 1$\sigma$ error. We proceed, therefore, to estimate the upper limits for the neutrino signal from unresolved gamma-ray blazars. To do this we follow the same procedure as for the sensitivity calculation, in which we substitute the null hypothesis given by the background simulations with the likelihood whose mean value is equivalent to the measured ${\rm C_P^{{\rm real}}}$ values as:

\begin{equation}
    {\rm \Delta_\kappa} = - 2 \left[ \log\mathcal{L_{S_\kappa \equiv {\rm C_P^{{\rm real}}}}}({\rm C_P^{{\rm real}}}) -  \log\mathcal{L^{{\rm S_\kappa}}}({\rm C_P^{S_\kappa}})\right]
\end{equation}

In Tab.~\ref{tab:sens} we report the 99\% confidence level sensitivity and upper limit.
Note that the ${\rm C_P^{{\rm real}}}$ in the first and last \fermilat{} energy bins are negative. As shown in the trials distributions in Fig.~\ref{fig:trialsdistrib}, fluctuation to negative values even in presence of expected positive correlation is possible and not rare. However it is important to comment that a negative 1-halo term could also be interpreted as an anti-correlation. 
% We believe this is not the case as anti-correlation is not supported by the cases of the middle \fermilat{} energy bins, in which we expect the most significant anisotropy signal from unresolved blazars according to \cite{UGRBaniso2018}. \\

\begin{table}[t]
    \centering
    \begin{tabular}{r|c|c}
     \toprule
    \multicolumn{3}{c}{Real data CAPS 1-halo term} \\
     \hline
    \fermi $\Delta$E &  ${\rm C_P}$ & $\delta{\rm C_P}$ \\
      ~[GeV] &  [(cm$^{-2}$s$^{-1}$sr$^{-1}$)~sr] & [(cm$^{-2}$s$^{-1}$sr$^{-1}$)~sr] \\
    \hline
      1-2    & -4.5E-13 & 4.6E-13\\
      2-5    & 8.5E-14  & 1.8E-13 \\
      5-10   & 2.6E-14  & 8.9E-14 \\
      10-25  & -3.5E-14 & 4.6E-14 \\
      \toprule
    \end{tabular}
    \caption{Results of the cross-correlation of real IceCube and \fermilat{} maps.}
    \label{tab:realCp}
\end{table}

% IClogE 	 FermiEbin 	 Cp 	 Cperr 
% 0.0 	 0 	 -4.476e-13 	 4.626e-13 
% 0.0 	 1 	 8.511e-14 	 1.848e-13 
% 0.0 	 2 	 2.568e-14 	 8.873e-14 
% 0.0 	 3 	 -3.555e-14 	 4.589e-14 

\begin{table}[t]
    \centering
    \begin{tabular}{r|c|c|c|c}
     \toprule
    \multicolumn{5}{c}{Sensitivity and upper limits summary table} \\
     \hline
    \fermi $\Delta$E &  \multicolumn{2}{c|}{Sensitivity} & \multicolumn{2}{c}{Limit} \\
    \hline
           ~    & $\kappa_{p-\gamma}$  & $\kappa_{p-p}$ & $\kappa_{p-\gamma}$  & $\kappa_{p-p}$ \\
    \hline
      1-2 GeV   & $<$2.80 & $<$5.59 & $<$0.80 & $<$1.59 \\
      2-5 GeV   & $<$0.73 & $<$1.45 & $<$0.84 & $<$1.67 \\
      5-10 GeV  & $<$0.50 & $<$1.00 & $<$0.58 & $<$1.15 \\
      10-25 GeV & $<$0.44 & $<$0.87 & {\bf $<$0.30} & {\bf $<$ 0.60} \\
      \toprule
    \end{tabular}
    \caption{Summary table of 99\% confidence levels sensitivities derived from simulations and upper limits computed from real data maps cross-correlation for \textit{p-p} and \textit{p-}$\gamma$ neutrino production channels. We highlight in bold the most stringent limits that we find.}
    \label{tab:sens}
\end{table}
% 99% Limits: [1.5927134892755583, 1.5203222030924688, 1.0977956274941758, 0.6281990195348748]
% 99% Sensitivities: [2.7833120555843713, 1.3222987197973457, 1.000270489599126, 0.8699842606768416]
% 99% Limits: [1.9184290364503283, 1.6685517410456678, 1.150067992706513, 0.599646404573492]
% 99% Sensitivities: [5.592322438517281, 1.4512212125906352, 1.000270489599126, 0.8699842606768416]

The results show that the measured ${\rm C_P}$ between the real \fermilat{} and IceCube data maps fall well within the range obtained from the simulations. This further validates the simulation procedure devised for this work. In the case of \textit{p-p} interactions, only in the highest gamma-ray energy bin considered we have an interesting upper limit with $\kappa<1$, which excludes at a C.L.$>$99\% that the totality of the gamma-ray emission from the contributing unresolved blazars is produced by neutral pions from \textit{p-p} (\textit{p-}$\gamma$) interactions. In all other energy bins we are not sensitive enough to make the same statement. In case of \textit{p-}$\gamma$ neutrino production channel, the sensitivities and upper limits are the same as for the \textit{p-p} interaction case but scaled by a factor of 2 (see Eq.~\ref{eq:pgamma}). The higher sensitivity of the cross-correlation at higher gamma-ray energies is attributable to the better angular resolution of the \fermilat{} data.

\begin{figure}[t]
    \centering
    \includegraphics[width=\textwidth]{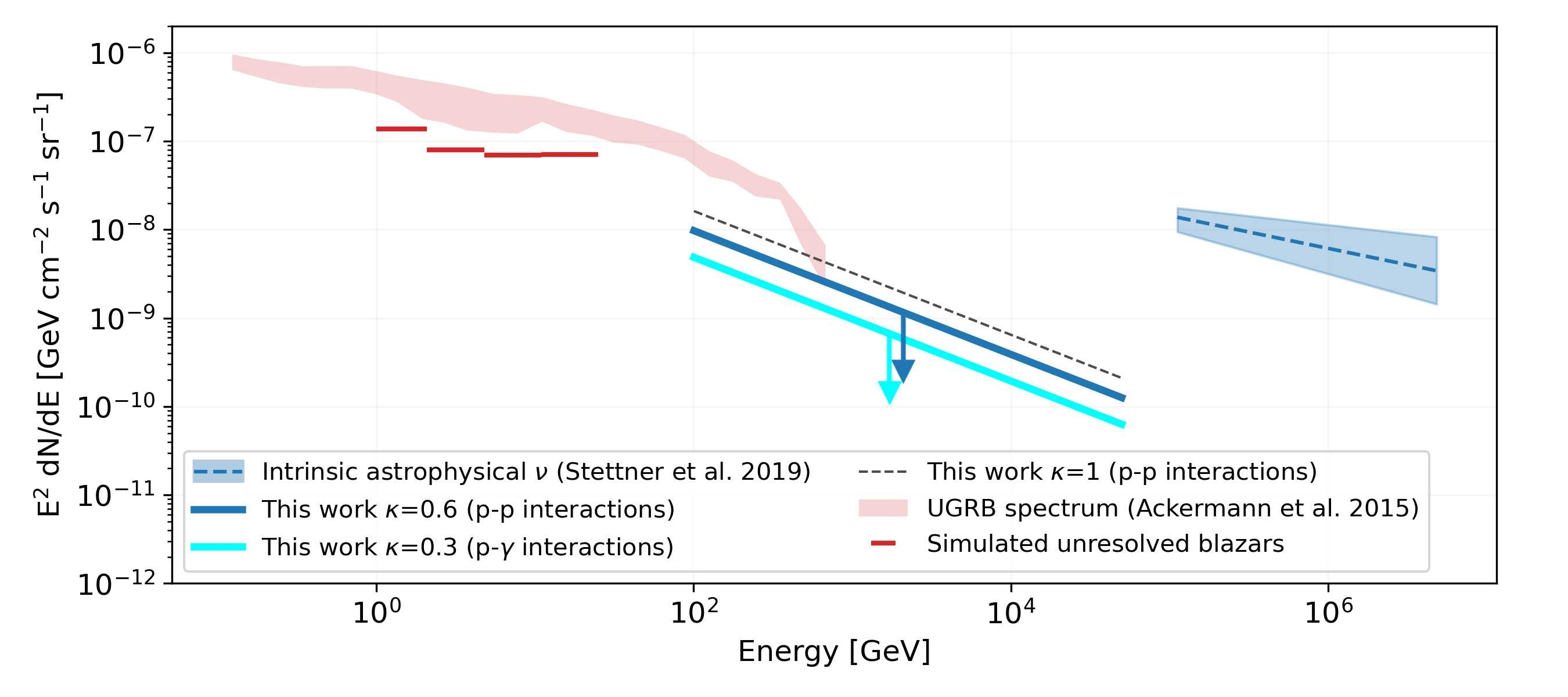}
    \caption{Contribution of the unresolved blazars that dominates the anisotropy of the UGRB to the astrophysical neutrino flux.}
    \label{fig:prospects}
\end{figure}

We can estimate the maximum contribution of the unresolved blazars that dominates the anisotropy measurement, by translating the upper limit on the parameter $\kappa$ into a constrain on the intrinsic neutrino flux to be compared to the one estimated by IceCube \citep{IClatestastro}. In order to do so, we derive the neutrino flux from the $\gamma$-ray flux using Eq.~\ref{eq:p-p} and Eq.~\ref{eq:pgamma} for \textit{p-p} and \textit{p-}$\gamma$ interactions, respectively, and then we apply the $\kappa$ factor of the most stringent limits (in bold in Tab.~\ref{tab:sens}). As outlined in Section~\ref{sec:sim}, we consider the $\gamma$-ray intrinsic spectra of each unresolved blazar to be describe by Eq.~\ref{eq:simspec}. At IceCube energies (above 100 GeV) this spectrum is a simple power law with index of -2.8, and we estimate that the total $\gamma$-ray differential flux at 100 TeV from all the unresolved simulated blazars is about $6.8\times 10^{-21}$ $\gamma$/cm$^{2}$/s/sr/GeV. This corresponds, according to Eq.~\ref{eq:p-p}, to $1.4\times 10^{-20}$ $\nu$/cm$^{2}$/s/sr/GeV. Because we are not using energy information in discriminating between signal and background events, our sensitivity is driven by the number of events observed from our assumed spectrum. Using simulations, we estimate 99\% of signal events with our spectrum fall between 100 GeV and 50 TeV, and therefore we report the constraints in this energy range. At 99\% C.L. we exclude that the unresolved blazars contribute to the astrophysical neutrino flux more than O(10\%) at 1 TeV, O(1\%) at 100 TeV.
This result is illustrated in Fig.~\ref{fig:prospects} and, together with the one obtained from the resolved blazars in the 1FLE catalog \citep{IceCube:2022zbd}, suggests that the blazar population, with brightness down to about an order of magnitude below the detection threshold of the \fermilat, can contribute to the astrophysical neutrino flux up to a couple of percent at about 100 TeV. We stress that any intrinsic energy cutoff and/or any additional components in the blazars spectra above measured TeV energies would not be included in our simulations. Generally, a cutoff would result in a weakening of the quoted limits, while an additional harder component would make the limit more stringent. See the first two figures in Appendix \ref{app:B} for further details.
% Central 90% of simulated energies is 110 GeV to 11000 GeV = 11 TeV
% Central 99% of the simulated energies is 100 GeV to 10**4.7 GeV = 50 TeV.

As a final consideration, we estimate how the sensitivity of this analysis will evolve for future studies. In particular, we consider 20 years of IceCube statistics, roughly when the IceCube/Gen2 \citep{ICGen2} configuration is expected to be in operation. We first assume no improvement in the angular resolution of the instrument, so we see the improvement due to the increased statistics only. Then we also report the ideal scenario of a perfect arrival direction reconstruction: this case can be seen as the lower limit of this sensitivity study. It is worth noting that in this ``ideal'' case, we are still limited by the LAT PSF and the presence of shot noise, which represent the lower limit to the predicted sensitivity. This is shown if Fig.~\ref{fig:sens} (right panel).

In this scenario we do not assume the UGRB signal to change with respect to the current set up: the \fermilat{} will probably not be operating at the time of IceCube/Gen2, and if it was we would have a more complete catalog of resolved gamma-ray sources, which makes it difficult to predict how the unresolved component will evolve (e.g. what kind of source populations will be dominating).

\section{Conclusions}
\label{sec:concl}
In summary, in this work we investigated the possibility to measure the spatial cross-correlation between the IceCube neutrino events (public data release by \cite{IC10yr}) and the unresolved gamma-ray background as measured by the LAT.
We devised a simulation pipeline for both IceCube event map and the \fermilat{} UGRB intensity map, given a list of sources and their intrinsic energy spectra. In this work we produce simulations from a list of unresolved blazars, to access the possible neutrino signal from this population of gamma-ray dim blazars. 
% The code is generic enough to allow the realization of simulated IceCube event from any given list of sources with a given location and a spectrum defined above 100 GeV. Therefore, we plan on releasing the simulation code to generate the IceCube event maps given a list of source location and spectra.
% \textcolor{blue}{Add a phrase on the availability of the code by Michael.}

We test two assumptions to derive our sensitivity to a neutrino-UGRB cross-correlation signal: 1) neutrino are produced via \textit{p-p} interactions or 2) via \textit{p-}$\gamma$ interactions. We vary the amount of injected signal by scaling the expected neutrino flux given the gamma-ray flux from a simulated unresolved blazar population. Such scaling is encoded in the $\kappa$ parameter of Eq.~\ref{eq:kappa}.

We find that, in the former case, the sensitivity of this study becomes interesting (below $\kappa=1$) at gamma-ray energy above 10 GeV; while in the latter case the sensitivity is constraining starting from 2 GeV.%, dow to $\kappa<0.33$ in the 10--25 GeV energy bin.}

The cross-correlation 1-halo term computed considering the real data maps is compatible with zero within the 68\% C.L. We therefore can only derive upper limits, which show similar trend to that revealed by sensitivity study. For \textit{p-p} interactions, the most constraining upper limit is above 10 GeV with a $\kappa < 0.60$ at the 99\% C.L.; for \textit{p-}$\gamma$ interactions the most constraining upper limit goes down to $\kappa < 0.30$ at the 99\% C.L.
%\textcolor{red}{This translate into a contribution of the brighter unresolved blazars (about 30\% of the total unresolved gamma-ray emission) to the astrophysical neutrino flux ...??? Need Ke}
Our analysis suggests that under the assumption that no energy cutoff and/or additional harder component is present above \fermilat{} energies, and that all gamma-rays of the unresolved blazars are produced by neutral pions from \textit{p-p} (\textit{p-}$\gamma$) interactions, up to 60\% (30\%) of the population may contribute the diffuse neutrino background. 
% Recalling that the anisotropic component of the unresolved blazars contributes to $\sim$30\% of the total UGRB intensity, this suggests that up to $\sim$20\% (10\%) of the UGRB emission \textcolor{blue}{due to blazars} can contribute to the high-energy astrophysical neutrino flux and still not be detected by cross-correlation analyses such as the one presented here. In terms of contribution to the astrophysical neutrino flux as observed by IceCube, 
We estimate a contribution to the neutrino flux of the order of the percent at 100 TeV (10 percent at 1 TeV) from the unresolved blazars that dominate the UGRB anisotropy spectrum as measured by the LAT.

Recently, a work by the IceCube Collaboration, reported a $\sim$4$\sigma$ excess probability to have a neutrino source in the direction of the nearby starburst Seyfert galaxy NGC~1068 \citep{ICM77}. This evidence suggests that AGNs of non-blazar type could contribute to the astrophysical neutrino flux. NGC~1068, being nearby, is detected by the LAT. However, misaligned AGN are generally faint in the GeV regime, and, being much more numerous than blazars, contribute to the total UGRB intensity roughly at the same extent as blazars \citep{IGRBmAGN}. A similar study as the one presented here, therefore, can be attempted for misaligned AGNs and will be subject of a future investigation. Furthermore, other smoother UGRB components, such as star forming galaxies, could also contribute and still be consistent with the observed null cross-correlation signal.

Finally, we estimated the projected sensitivity at 20 years of IceCube data. Assuming the current angular resolution, the sensitivity improves to $\kappa\sim$0.45 (0.23) for \textit{p-p} (\textit{p-}$\gamma$) interaction. Such improvement is a factor of about 30\% with respect to the sensitivity at 6 years, and therefore it will be worth attempting such a study again in the future with an enhanced neutrino statistics.

% Assuming an optimistic improvement in direction reconstruction for IceCube future configurations (i.e. IceCube/Gen2) of about 20\%, and a reprocessing of the past data, the sensitivity does not significantly improve further. 

\bigskip
\bigskip

\section*{Acknowledgments}
We acknowledge Micheal Larson and Ke Feng for crucial insights on the correct use of the IceCube public data and on the high-energy neutrino astrophysics. We also acknowledge Alessandro Cuoco for serving as LAT internal referee and for the useful suggestions on how to approach the cross-correlation analysis. MN and MC acknowledge that the material is based upon work supported by NASA under award number 80GSFC21M0002. 
% The work of KF is supported by the Office of the Vice Chancellor for Research and Graduate Education at the University of Wisconsin-Madison with funding from the Wisconsin Alumni Research Foundation. KF acknowledges support from NASA (NNH19ZDA001N-Fermi, NNH20ZDA001N-Fermi) and National Science Foundation (PHY-2110821).  
LM acknowledges that support for this work was provided by NASA through the NASA Hubble Fellowship grant \#HST-HF2-51486.001-A, awarded by the Space Telescope Science Institute, which is operated by the Association of Universities for Research in Astronomy, Inc., for NASA, under contract NAS5-26555.
% \textcolor{red}{Anybody else wants to acknowledge?} 
Some of the results in this paper have been derived using the healpy and HEALPix package.
The \textit{Fermi} LAT Collaboration acknowledges generous ongoing support
from a number of agencies and institutes that have supported both the
development and the operation of the LAT as well as scientific data analysis.
These include the National Aeronautics and Space Administration and the
Department of Energy in the United States, the Commissariat \`a l'Energie Atomique
and the Centre National de la Recherche Scientifique / Institut National de Physique
Nucl\'eaire et de Physique des Particules in France, the Agenzia Spaziale Italiana
and the Istituto Nazionale di Fisica Nucleare in Italy, the Ministry of Education,
Culture, Sports, Science and Technology (MEXT), High Energy Accelerator Research
Organization (KEK) and Japan Aerospace Exploration Agency (JAXA) in Japan, and
the K.~A.~Wallenberg Foundation, the Swedish Research Council and the
Swedish National Space Board in Sweden. Additional support for science analysis during the operations phase is gratefully acknowledged from the Istituto Nazionale di Astrofisica in Italy and the Centre National d'\'Etudes Spatiales in France. This work performed in part under DOE Contract DE-AC02-76SF00515.

\bibliography{sample631}{}
\bibliographystyle{aasjournal}

\appendix

\section{IceCube PSF and CAPS correction}
\label{app:A}
\begin{figure}[h]
    \centering
    \includegraphics[scale=0.35]{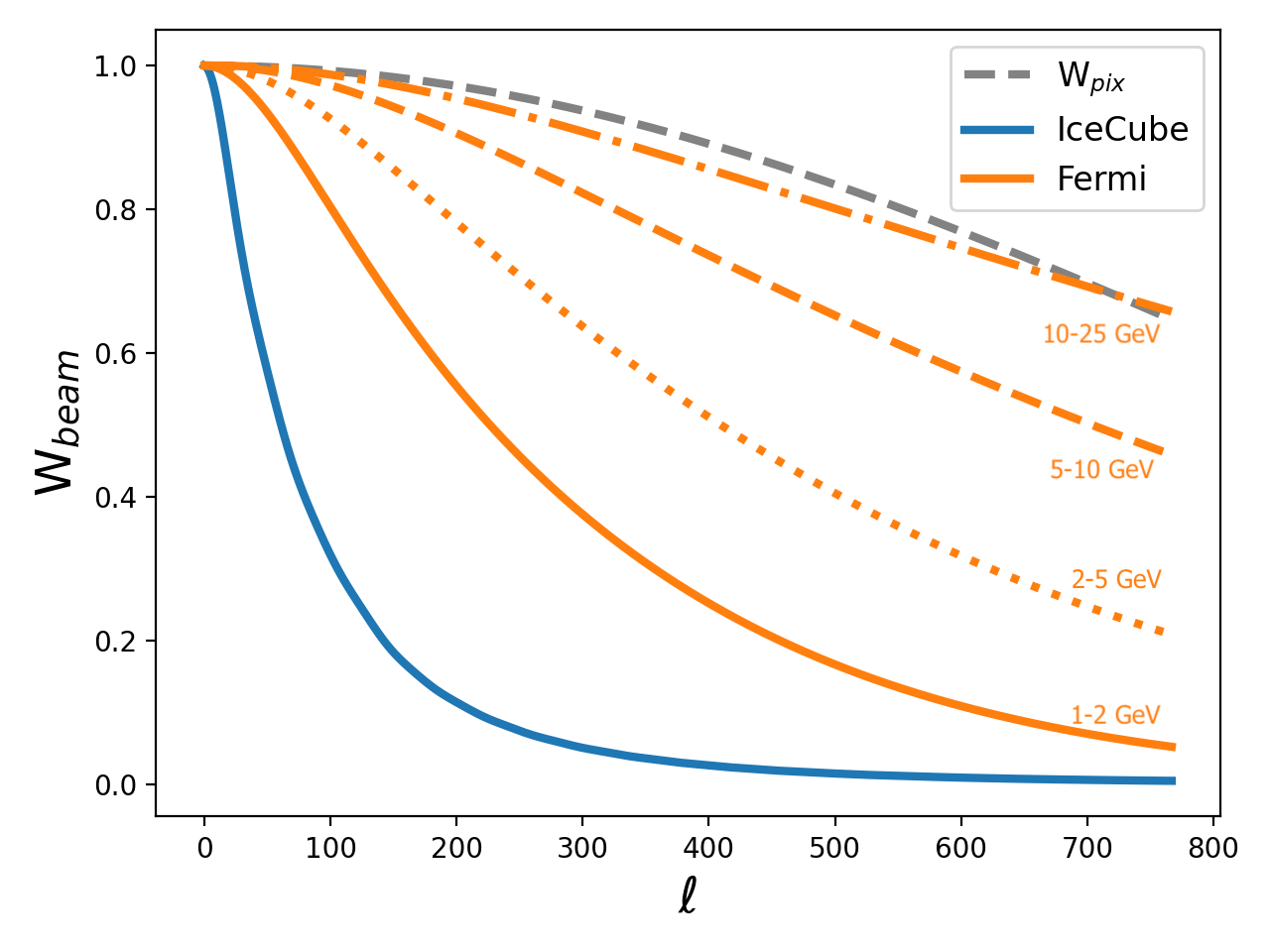}
    \includegraphics[scale=0.35]{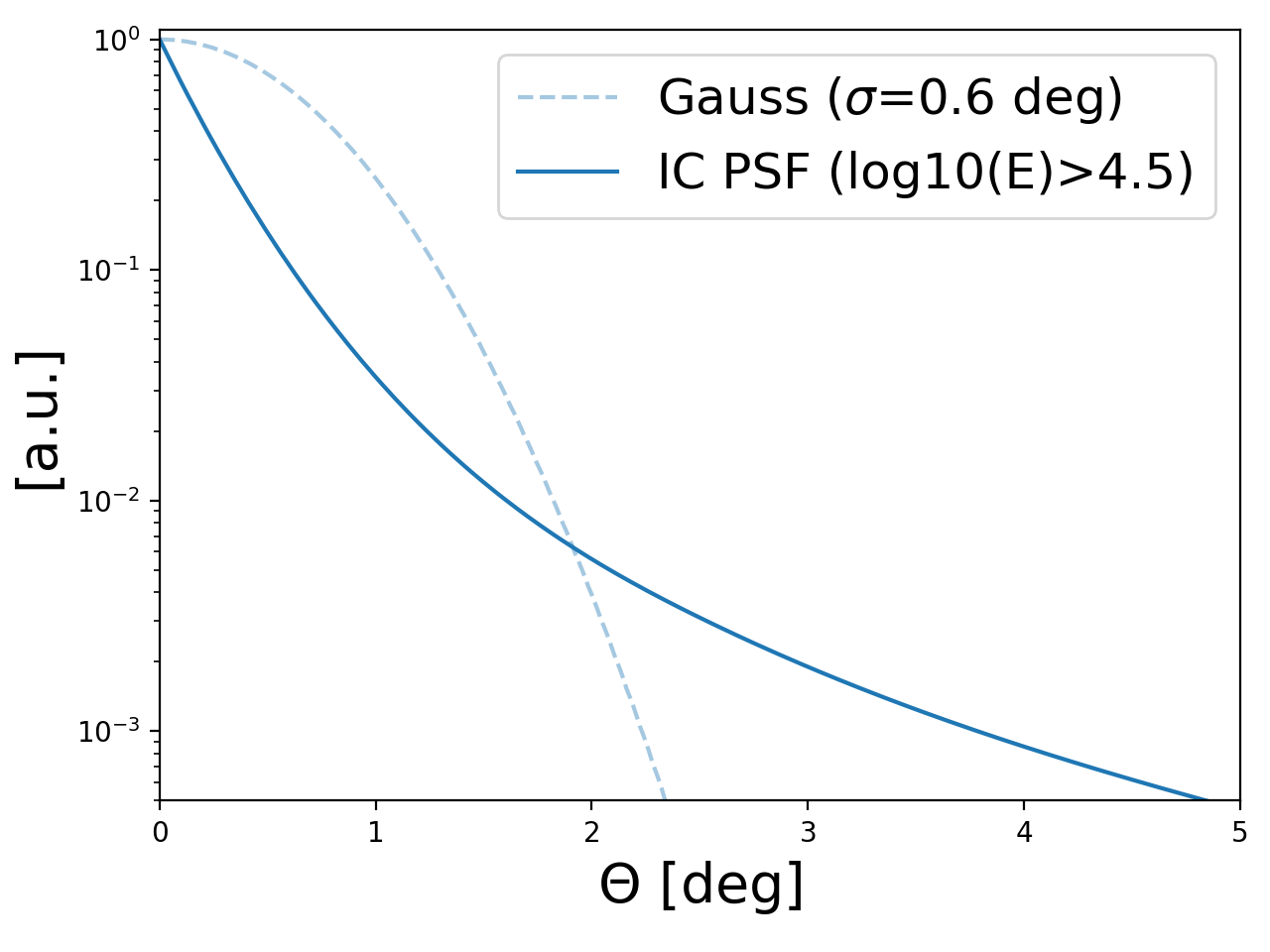}
    \includegraphics[scale=0.35]{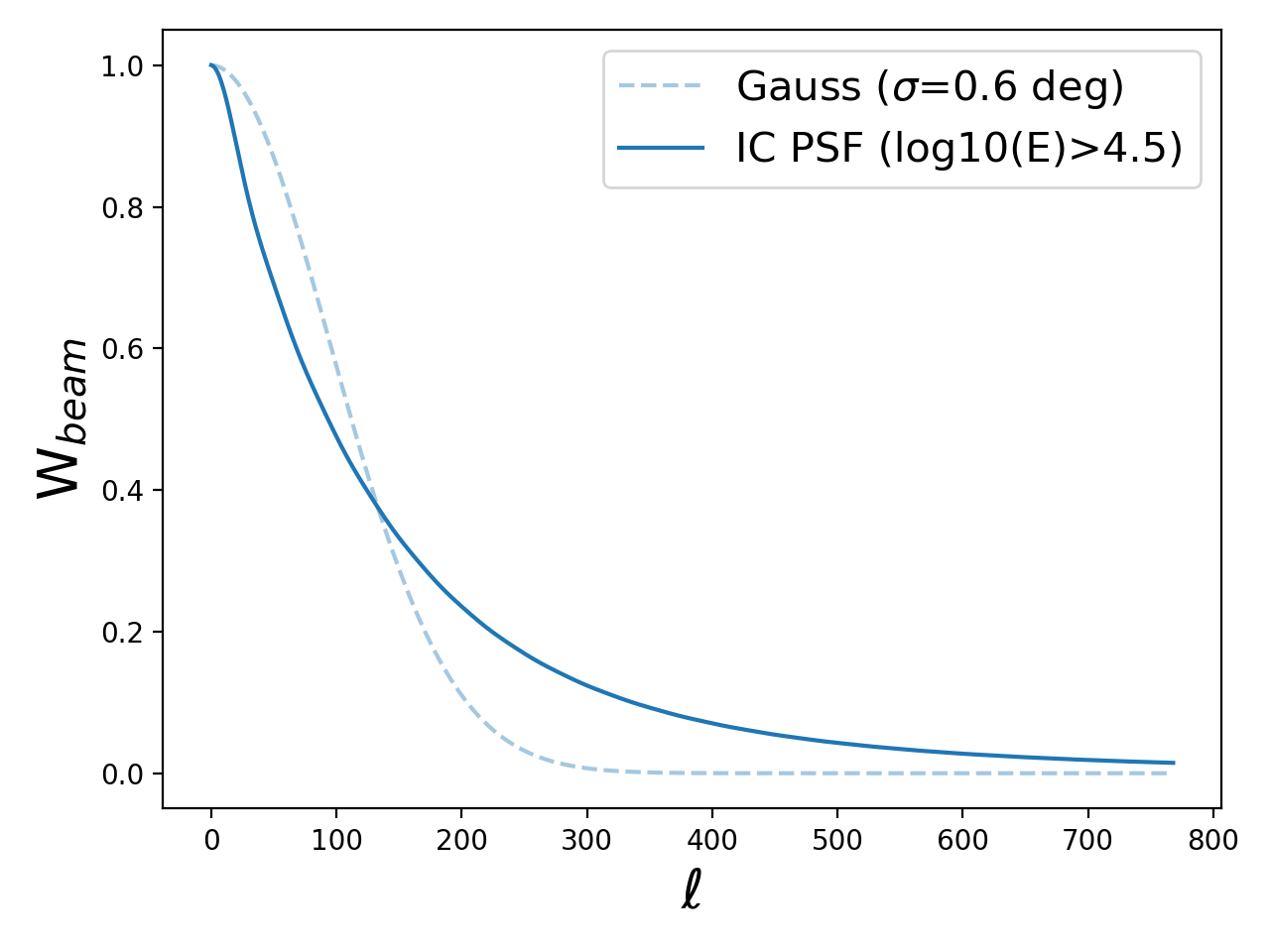}
    \caption{Left: Window beam functions applied to the maps in order to correct for the instrumental angular resolution. In orange we show the correction functions for the four \fermilat{}  energy bins, while in blue is the correction for the IceCube map. In gray we show the pixel window function used to account for the finite pixeling of the maps. Middle: Comparison between the PSF profile from the public data release \citep{IC10yr} and a Gaussian profile with $\sigma=0.6$ deg. Right: Comparison between the beam window functions derived from the PSF and Gaussian profiles shown in the middle panel.}
    \label{fig:Wbeams}
\end{figure}

We derive the IceCube PSF profile for our selected samples from the smearing matrices provided with the official data release \cite{IC10yr}, considering the response functions for year 2012-2018 configuration ({\tt IC86\_II\_effectiveArea}), and selecting the Northern hemisphere only (declination bins above $-5^\circ$). 
The response functions are weighted by the product of IceCube's effective area and an unbroken $E^{-2}$ power law in order to properly account for the relative contributions from high and low energy events.
The resulting PSFs are fit with a univariate spline \cite{2020SciPy-NMeth}, resulting in a smooth description of IceCube's pointing averaged over the assumed power law spectrum. 
An example of the obtained PSF is shown in Fig.~\ref{fig:Wbeams}.
% \textcolor{blue}{Michael: add more details?}.

We compute the window functions from the obtained PSF profile using Eq.~\ref{eq:Wbeam} and the result is shown in the left plot of Fig.~\ref{fig:Wbeams} (blue line): as expected the plot shows how the correction due to IceCube angular resolution is significantly more important that for \fermilat's. Therefore, we want to emphasize the importance of deriving the correct PSF profile from IceCube data, with respect to adopt the Gaussian approximation. In the middle and left plots of Fig.~\ref{fig:Wbeams}, we compare the Gaussian vs data-driven profiles of the IceCube PSF and the derived $W_{beam}$ respectively. At the small angular scales (large multipoles) we are interested in, the Gaussian profile is clearly mis-representing the true angular smearing of the IceCube data.

\section{Additional plots and discussion}
\label{app:B}
\begin{figure}[htb]
    \centering
    \includegraphics[scale=0.28]{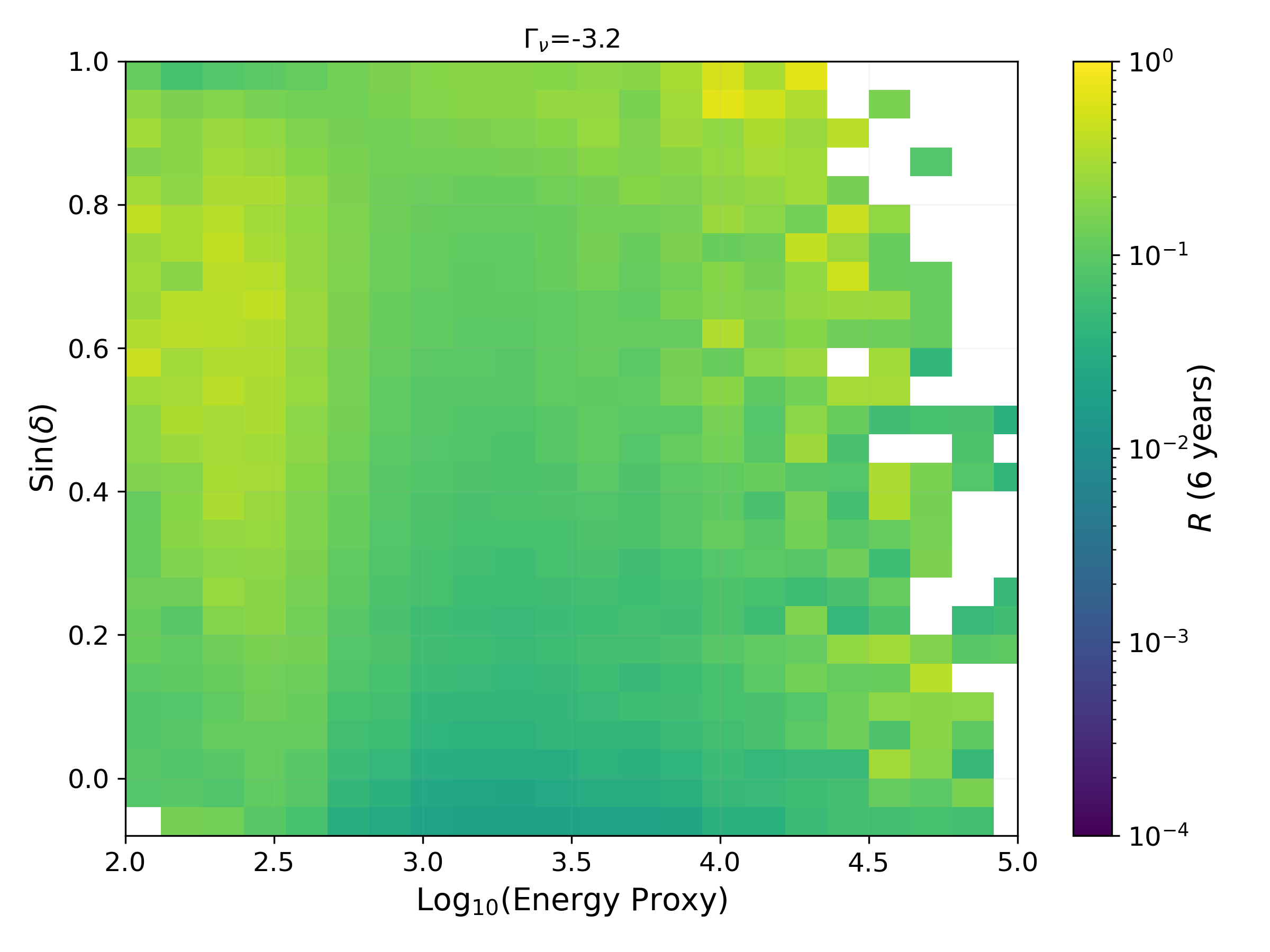}
    \includegraphics[scale=0.28]{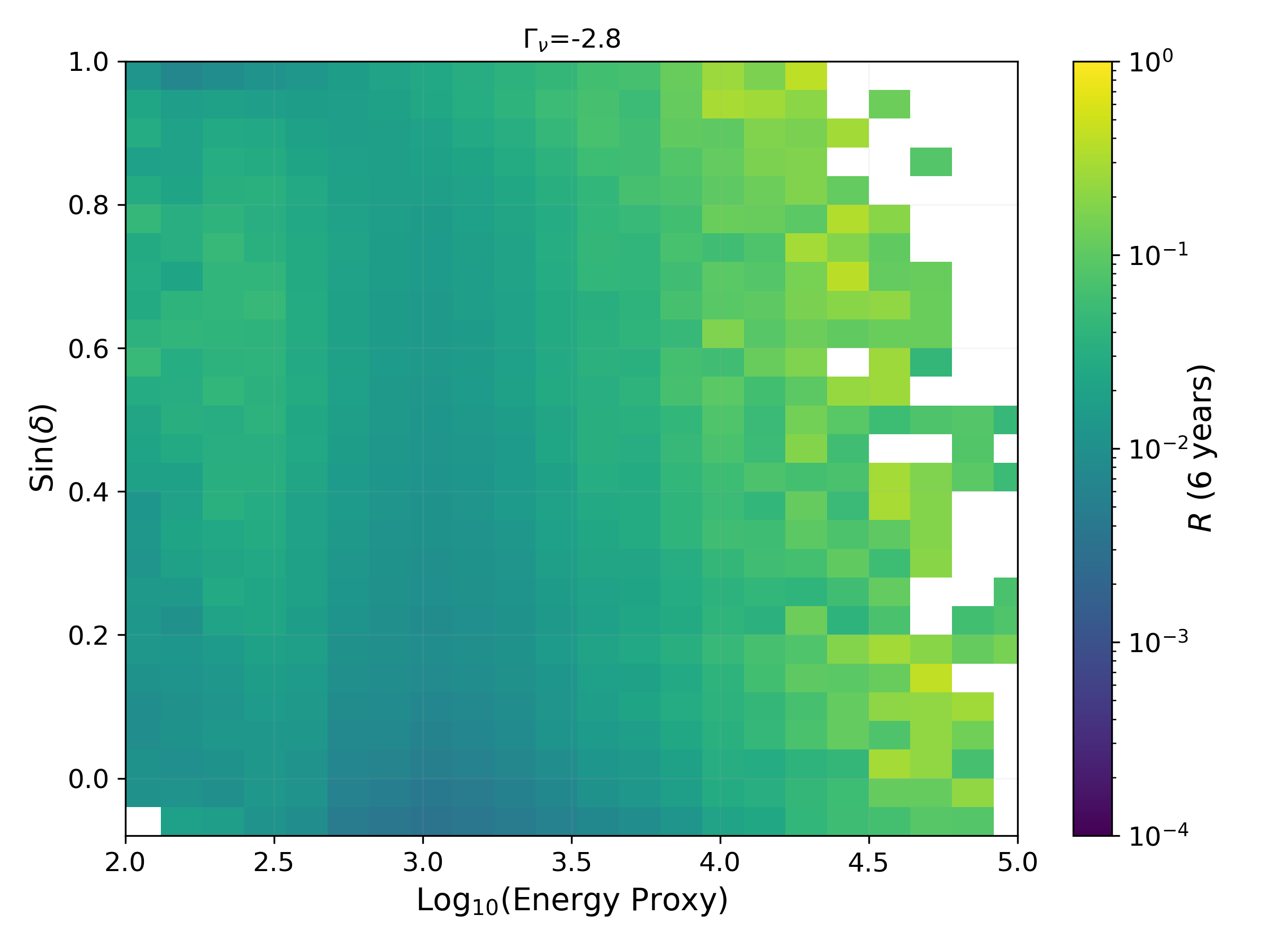}
    \includegraphics[scale=0.28]{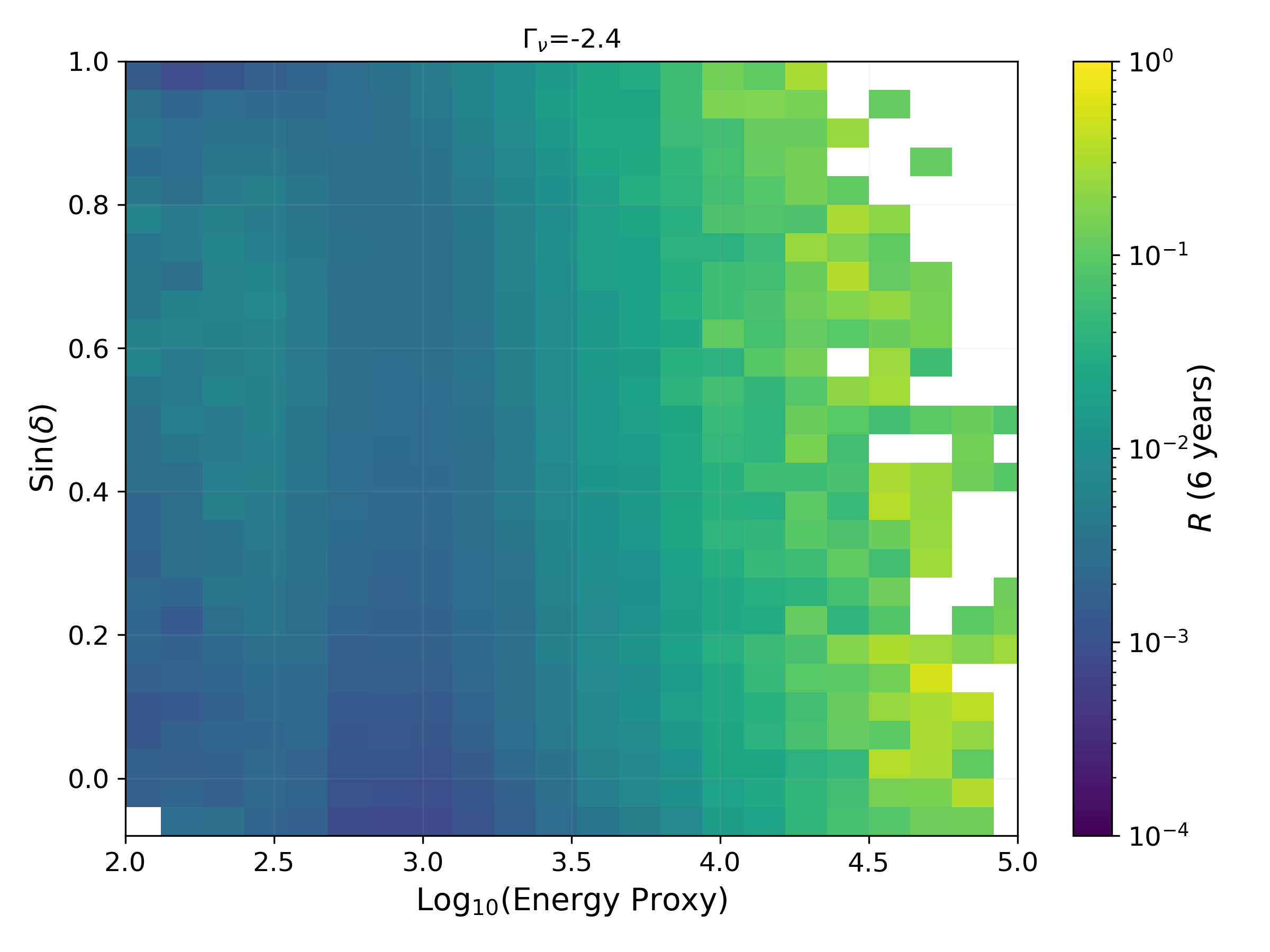}
    \caption{\textcolor{blue}{Variation of IceCube sensitivity in the Northern hemisphere assuming different spectral indices of the power-law spectrum for the astrophysical signal (the normalization is kept the same as in \cite{Astronuflux}). The plots are generated using the same method described in Section \ref{subsec:mask} assuming different spectral indices ($\Gamma_\nu=-3.2, -2.8, -2.4$ for the left, middle, and right panels, respectively). Energy Proxy is in units of GeV.}}
    \label{fig:sensVSindex_2d}
\end{figure}

\begin{figure}[htb]
    \centering
    \includegraphics[scale=0.4]{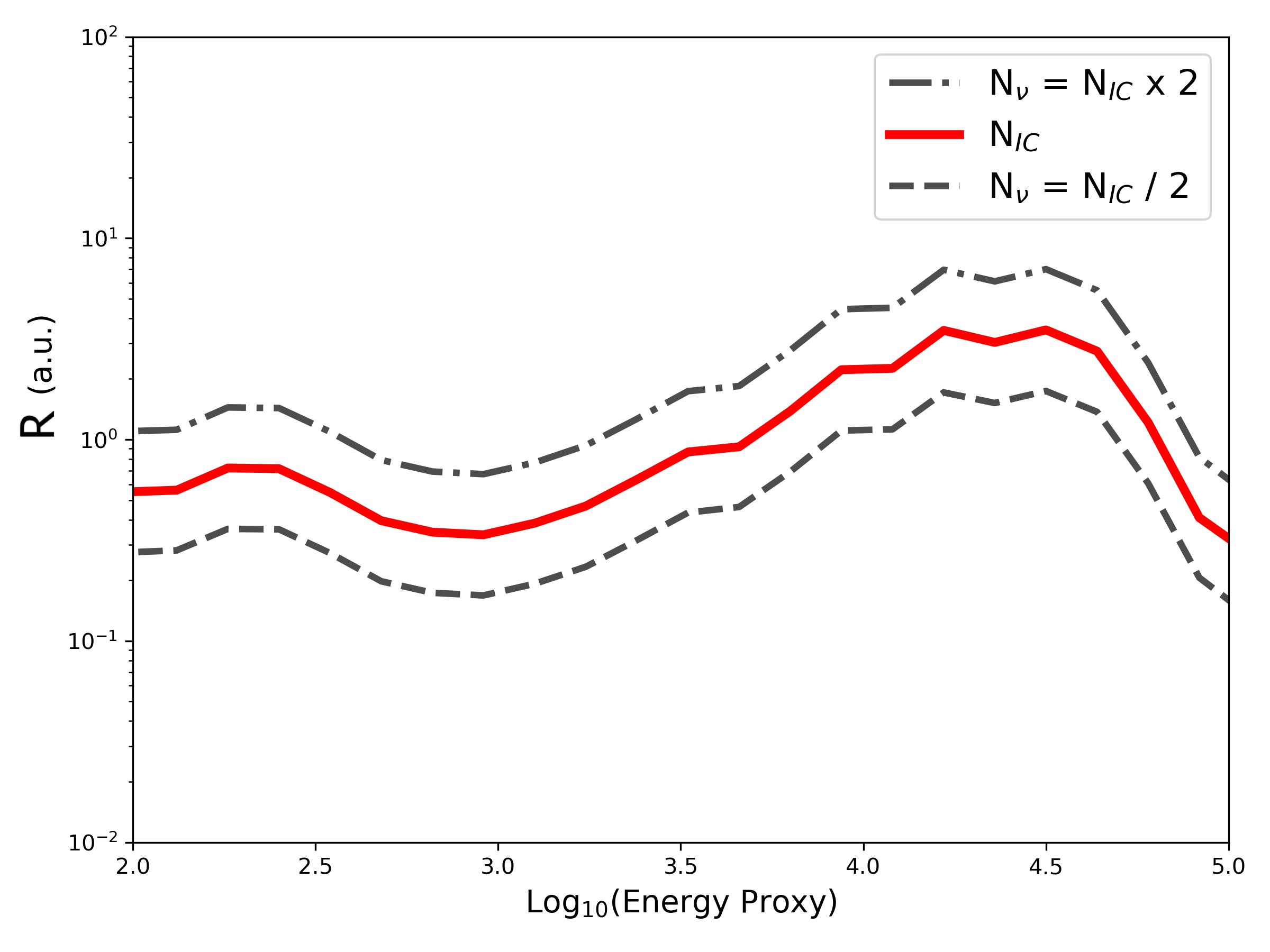}
    \includegraphics[scale=0.4]{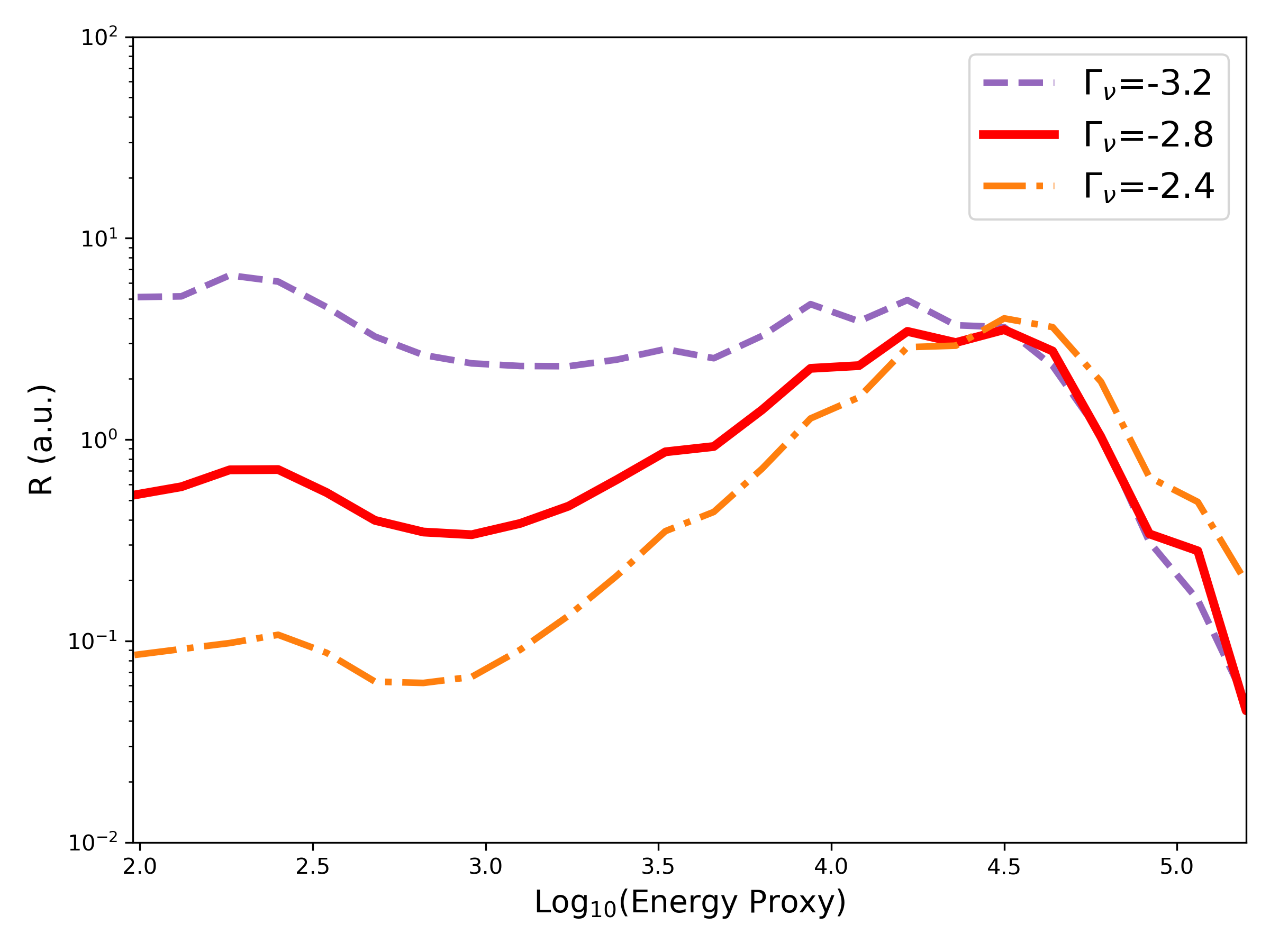}
    \caption{Variation of IceCube sensitivity as a function of the energy for the Northern hemisphere, for the case of a variation in normalization (left) and the spectral index (right) of the power law. The nominal normalization is $N_{IC}=1.44\times10^{-18}{\rm GeV}^{-1}{\rm cm}^{-2}{\rm s}^{-1}{\rm sr}^{-1}$ at 100 TeV as measured by \citep{IClatestastro}. Energy Proxy is in units of GeV.}
    \label{fig:sensVSindex}
\end{figure}

\begin{figure}[htb]
    \centering
    \includegraphics[scale=1]{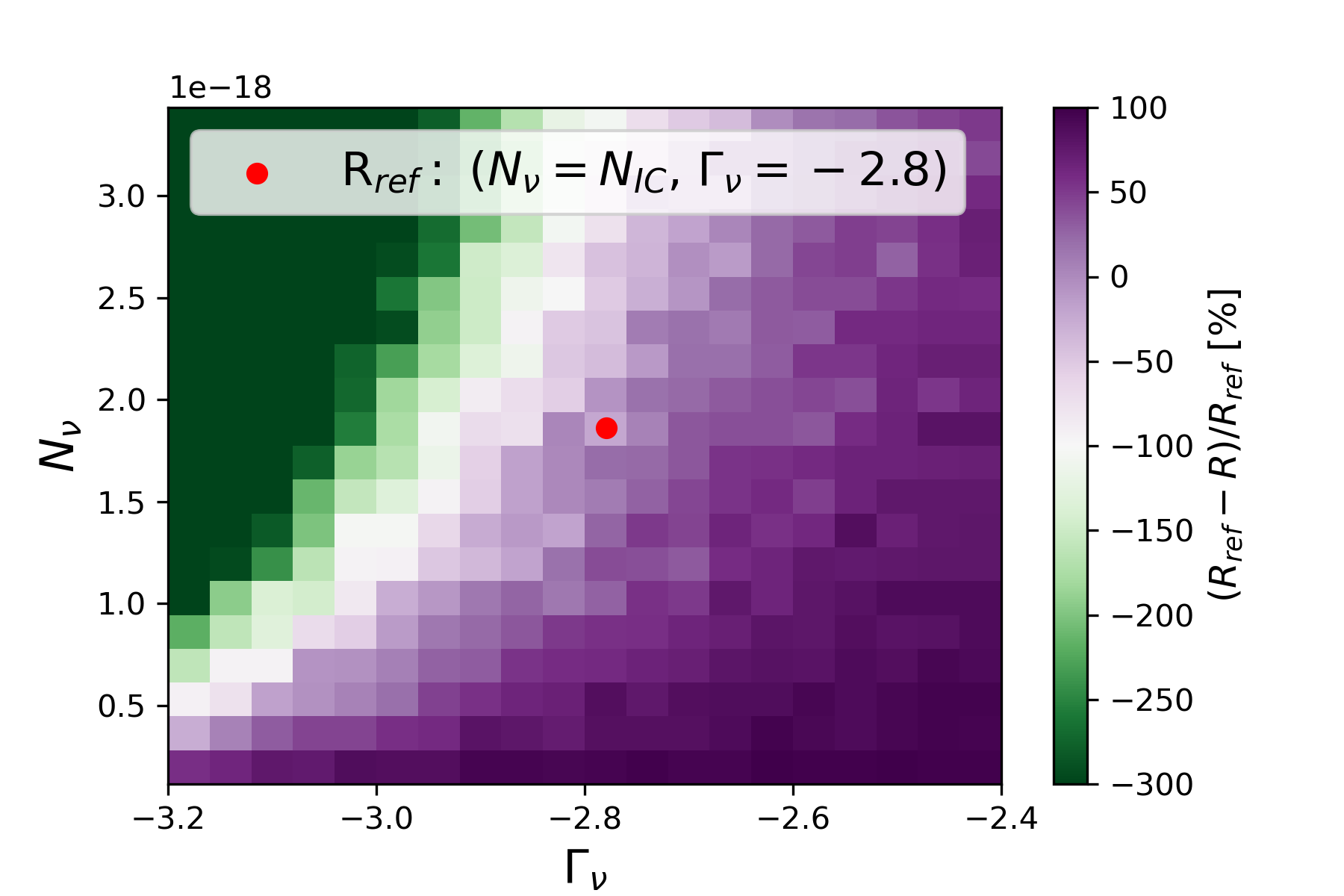}
    \caption{Variation of the parameter $R$ with respect to nominal values of assumed power-law normalization and index. Negative variations indicate better sensitivity, while positive variations indicate worse sensitivity.. $N_{IC}=1.44\times10^{-18}{\rm GeV}^{-1}{\rm cm}^{-2}{\rm s}^{-1}{\rm sr}^{-1}$ at 100 TeV as measured by \citep{IClatestastro}}
    \label{fig:newSens}
\end{figure}

In this section we report some additional plots and discussions to complement the study described in the main text. In particular, we discuss how variations in the assumed extrapolation of the gamma-ray flux to IceCube energies affects the sensitivity. In the text we mentioned that the simple extrapolation of the gamma-ray spectrum from \citep{LeaSimBlazar} to determine the expected neutrino events does not account for possible spectral breaks, cutoffs and/or additional components. However, simulating the complete range of possible scenarios is computationally infeasible. Nevertheless, we can still study the variation of the sensitivity by noting that our sensitivity is proportional to the total number of signal counts in the IceCube map. This number varies with the parameters of power-law spectrum assumed. Ignoring the absolute values of the sensitivity, its variations with the power-law parameters for the neutrino signal can be assessed using the IceCube response functions. We follow the same procedure described in Section~\ref{subsec:mask} to study the best sky hemisphere for astrophysical neutrino signal. In this framework, the intrinsic astrophysical neutrino spectrum is assumed to be a power law $$\frac{d\Phi_\nu}{dE_\nu}\left(E_\nu\right)= N_\nu \left(\frac{E_\nu}{100~{\rm TeV}}\right)^{\Gamma_\nu}~,$$
where $N_\nu$ is in ${\rm GeV}^{-1}{\rm cm}^{-2}{\rm s}^{-1}{\rm sr}^{-1}$. We calculate the ratio, $R$, between the estimated signal counts over the observed counts, which gives us a proxy of IceCube sensitivity to the assumed astrophysical neutrino signal. In Fig.~\ref{fig:sensVSindex_2d} we show some examples of $R$ as a function of the declination (here we focus only on northern hemisphere) and the reconstructed energy. The three panels show three different assumptions of the power-law index $\Gamma_\nu=-3.2, -2.8, -2.4$, while the normalization is kept the same. Note that the central value corresponds to the $\delta_2$ parameter in Eq.~\ref{eq:simspec}, which in our analysis is extrapolated to IceCube energies to build the simulated neutrino maps.
From these 2D sensitivity maps, we can study the variations with energy by integrating over the declination, deriving the curves in Fig.~\ref{fig:sensVSindex}. These latter plots explicitly illustrate the effect of the different spectral normalization and indices of the assumed power law. In general, a change in the spectrum normalization (as investigated in this work by varying the parameter $\kappa$), translates in a rigid proportional shift of the $R$ up or down (Fig.~\ref{fig:sensVSindex}, left panel). On the other hand, if we kept the normalization fixed to the measured value \citep{Astronuflux}, harder (steeper) spectra slightly improve $R$ at higher (lower) energies, while reducing it at lower (higher) energies (Fig.~\ref{fig:sensVSindex}, right panel). 
In this work, however, we do not directly include neutrino energy information. Our sensitivity is only determined by the estimated counts rate of detected astrophysical neutrino events above 100~GeV. Hence, we derive a global sensitivity-proxy, $R$, defined as the ratio of total estimated signal counts over total observed counts in the northern hemisphere above 100 GeV. The variation with respect to nominal values of normalization and index of the assumed power law is illustrated in Fig.~\ref{fig:newSens}, where the nominal parameters are also reported in the legend. Generally speaking, the combination of parameters that would return a higher number of signal counts (higher normalization at 100 TeV with softer spectra) would be expected to improve the sensitivity to $\kappa$ (towards greener values in the plot), as opposed to lower normalization with harder spectra, which would worsen the sensitivity (purple values in the plot). However, the trend is not linear in spectral index, as shown by the white values in the plot. This band roughly corresponds to an improvement of a factor of 2 with respect to the nominal values. 
These tests assess the impact of an unknown spectral break or new component arising from the same sources producing the observed UGRB sky in \fermilat. The results demonstrate that the choice of how to model the flux of IceCube neutrino events from the observed UGRB can strongly influence the final results. Additional assumptions, such as non-power-law spectra, are also possible, but are beyond the scope of this work.
%, suggesting that an improvement in sensitivity for the presence of an additional component is not as big as the worsening due to the presence of an energy cutoff.
\\

\begin{figure}[htb]
    \centering
    \includegraphics[scale=0.4]{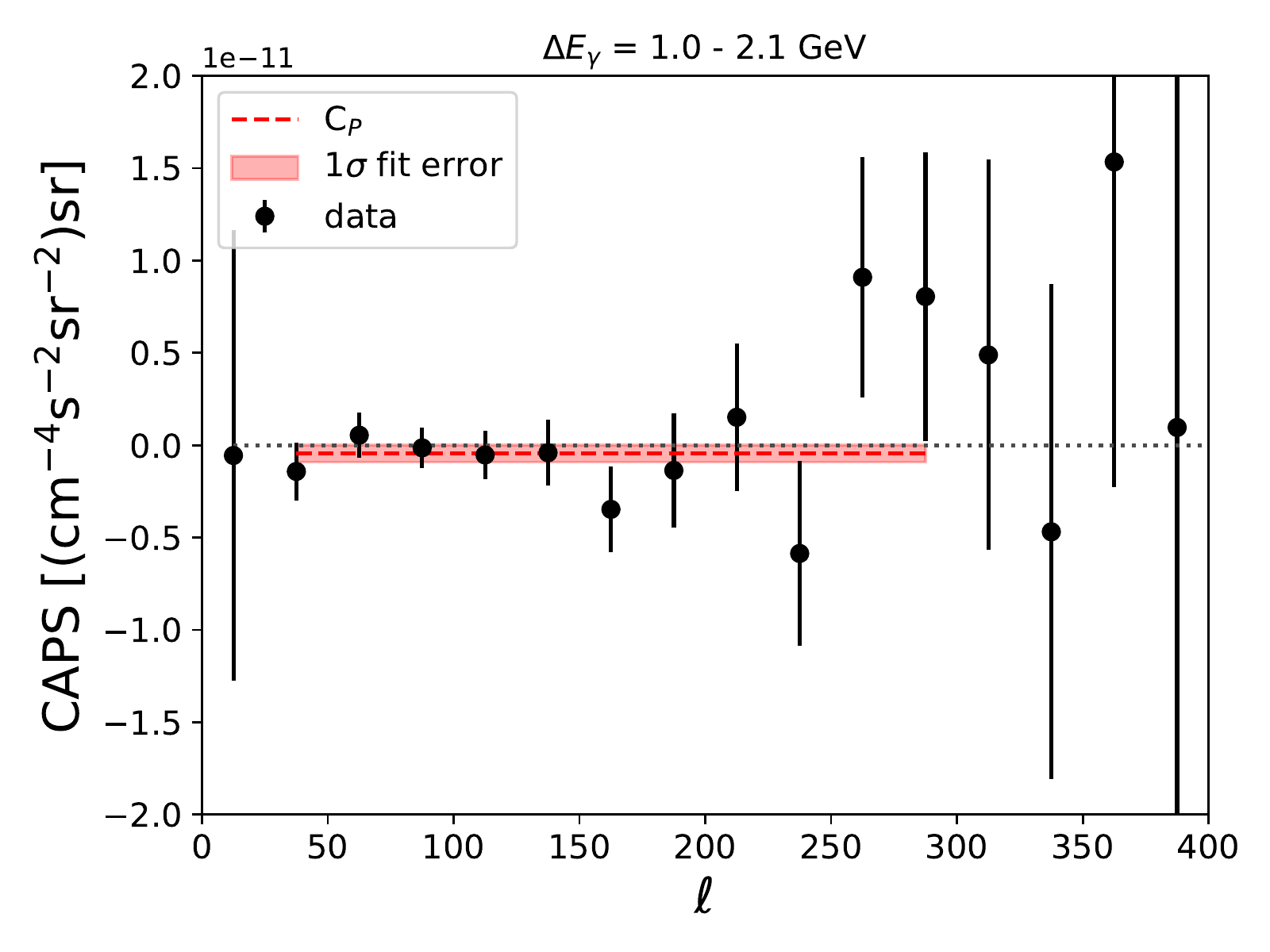}
    \includegraphics[scale=0.4]{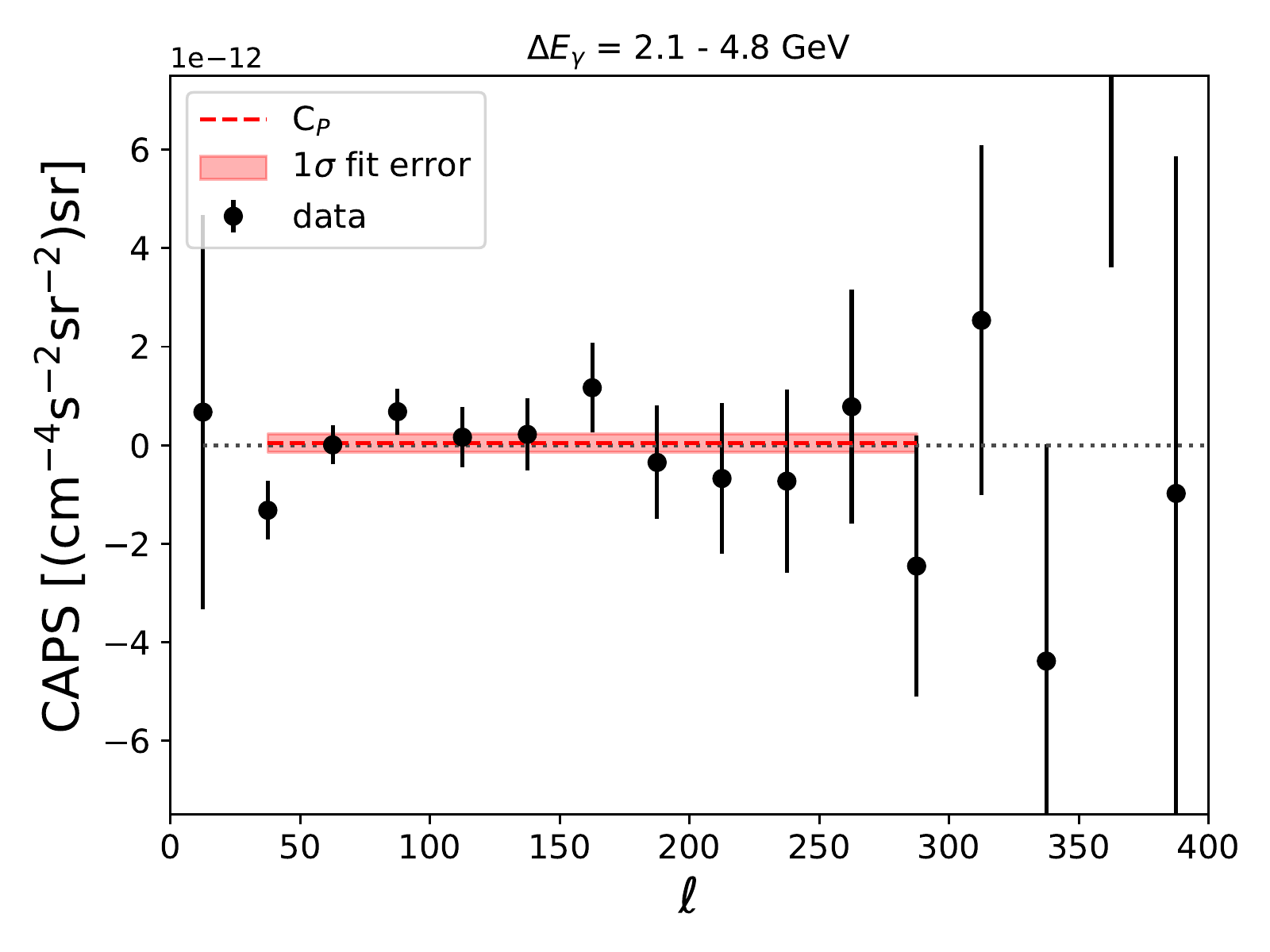}
    \includegraphics[scale=0.4]{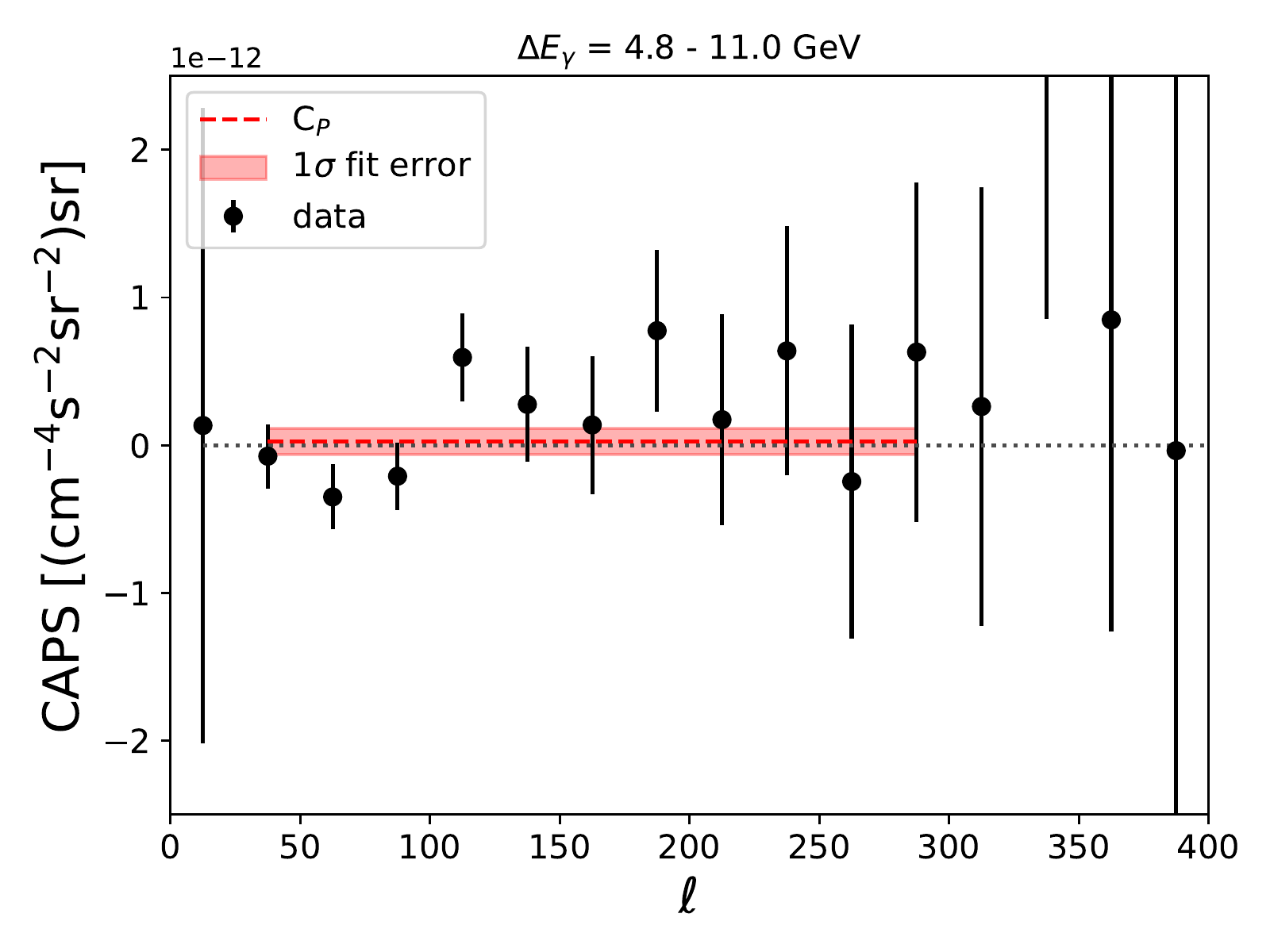}
    \includegraphics[scale=0.4]{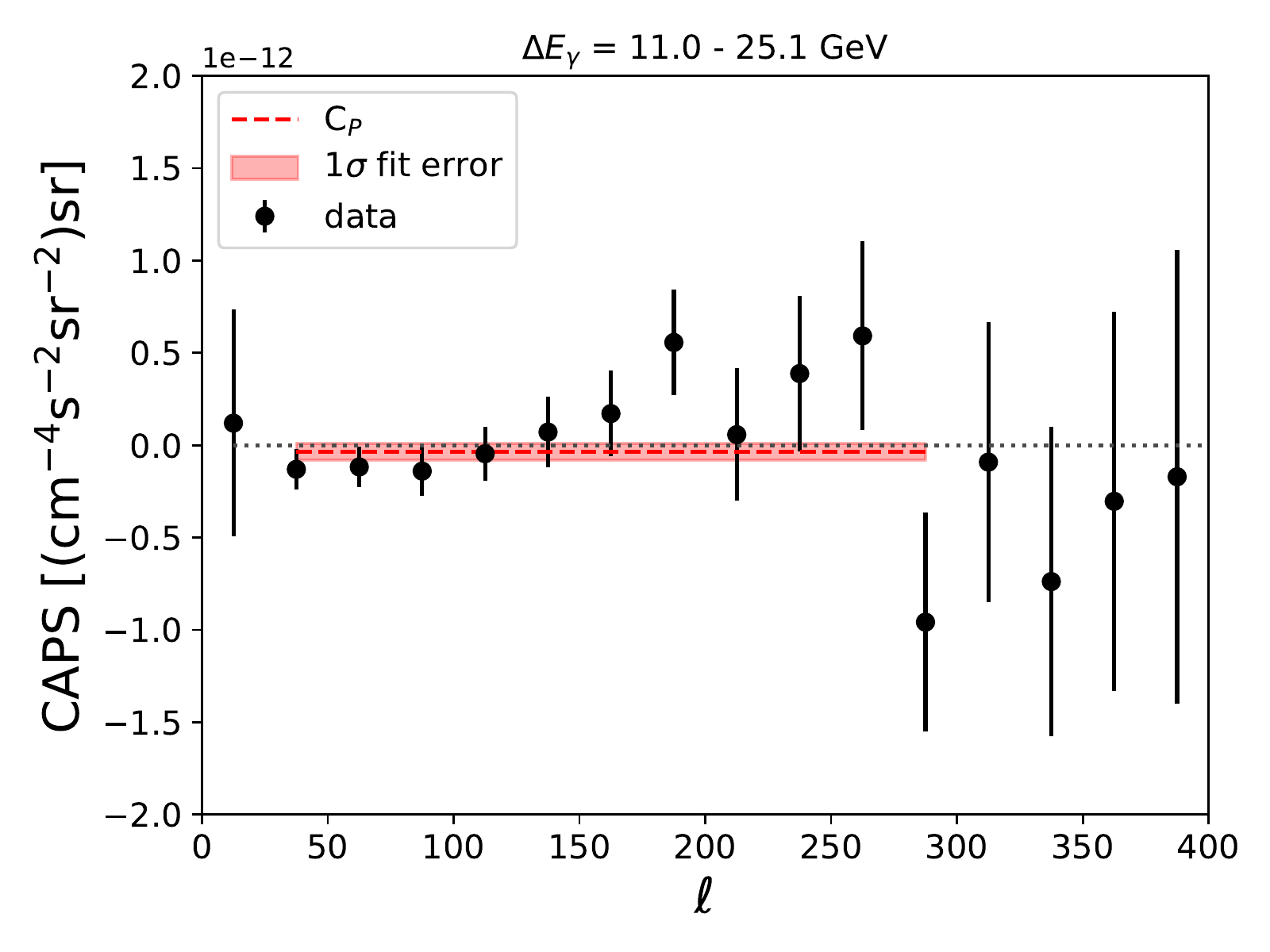}
    \caption{CAPS for the real \fermilat{} and IceCube maps.}.
    \label{fig:real_cross}
\end{figure}

In Fig.~\ref{fig:real_cross} we show the obtained CAPS for the cross-correlation of the real data maps. The four panels correspond to the four \fermilat energy bins, and the best-fit \cp{} is also shown with the one sigma uncertainty. 

\begin{figure}[htb]
    \centering
    \includegraphics[scale=0.3]{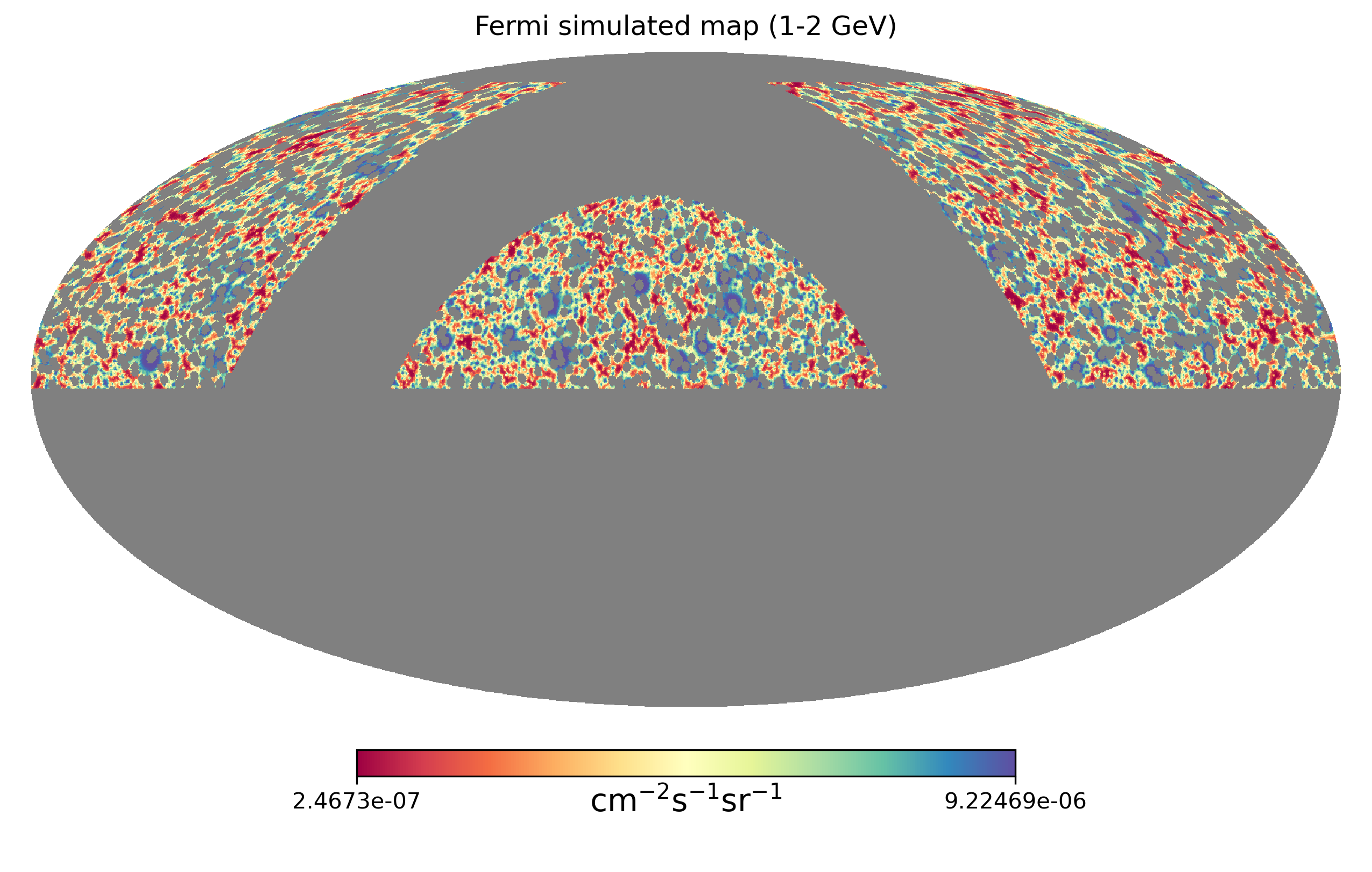}
    \includegraphics[scale=0.3]{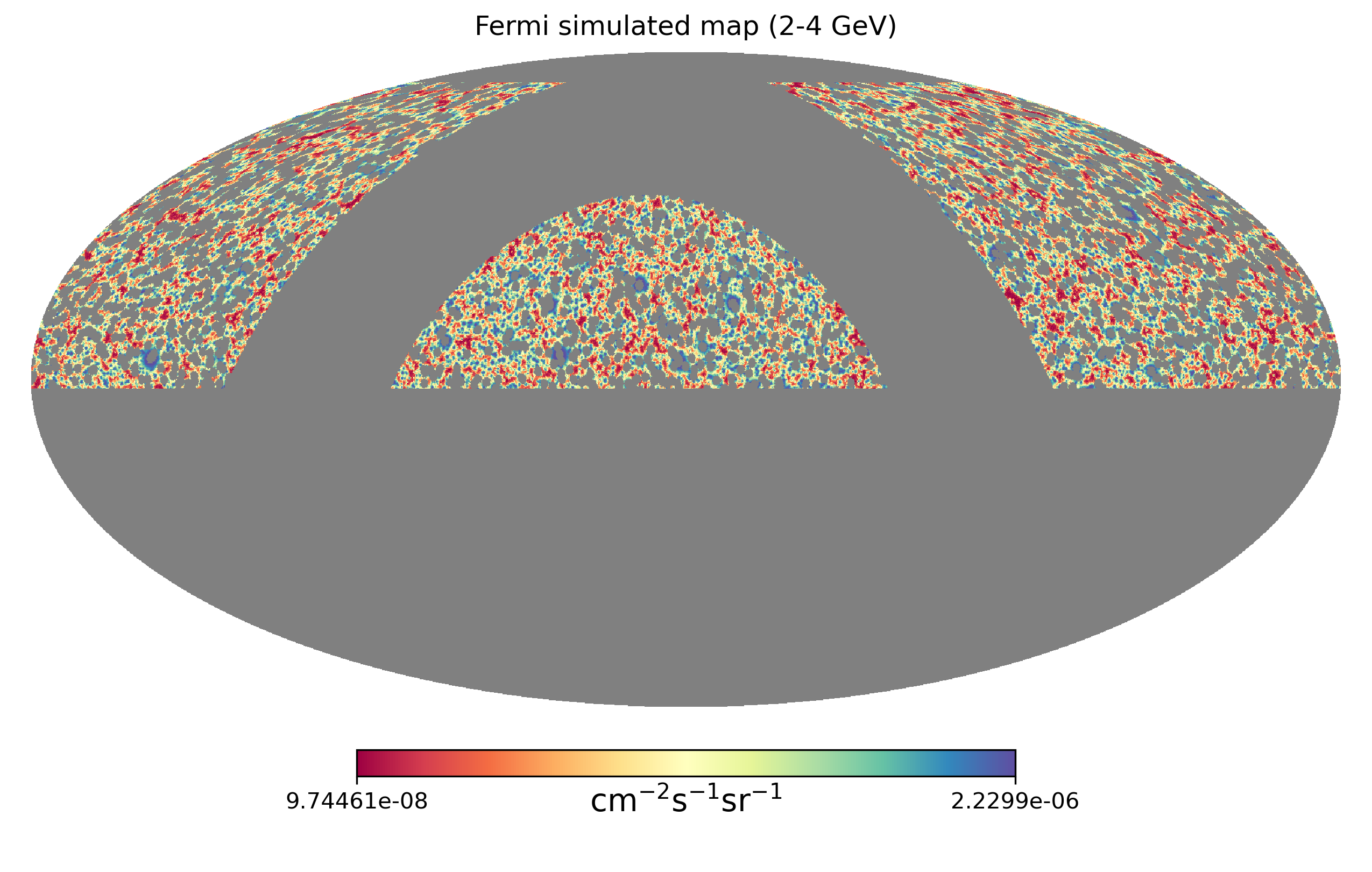}
    \includegraphics[scale=0.3]{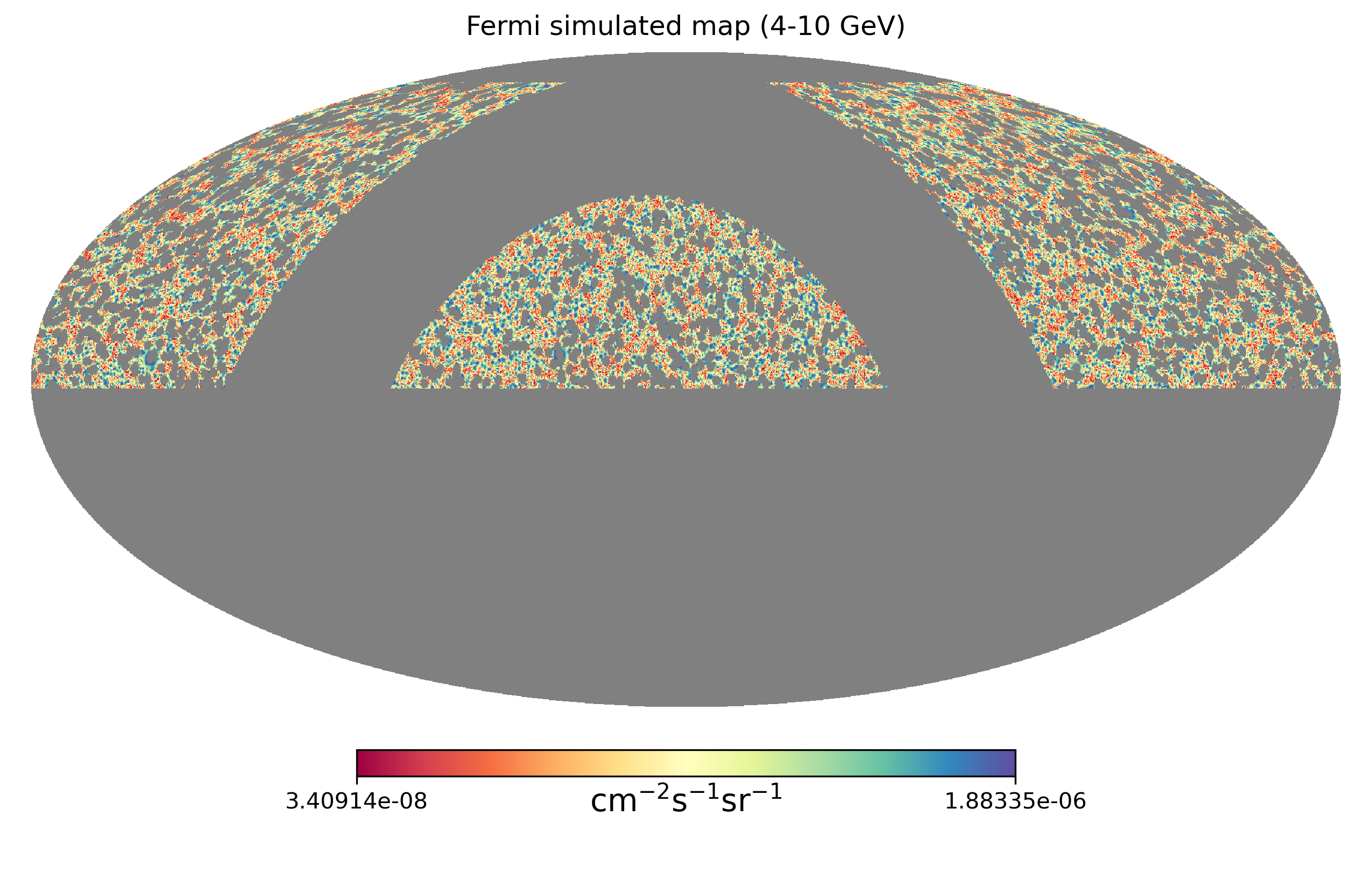}
    \includegraphics[scale=0.3]{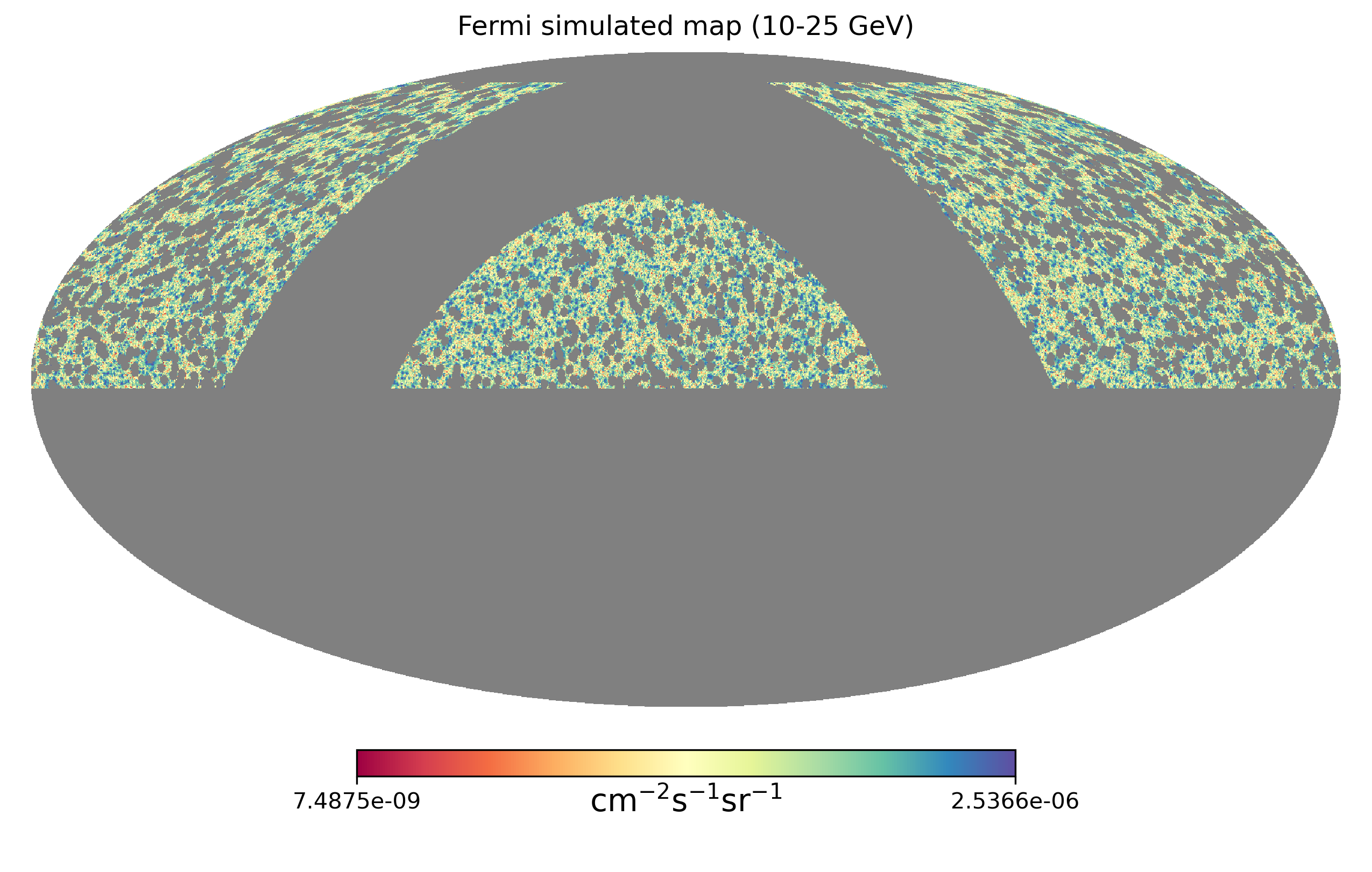}
    \caption{Simulated \fermilat{} UGRB maps for the four energy bins considered. The colorbar is in unit of flux cm$^{-2}$s$^{-1}$sr$^{-1}$.}.
    \label{fig:sim_fermi}
\end{figure}

In Fig.~\ref{fig:sim_fermi} we show the masked simulated Fermi-LAT{} UGRB maps.

\end{document}